\documentclass[epsfbox]{iopbk2e}  
\usepackage[english]{babel}
\usepackage{epsfig}
\def\simlt{\mathrel{\rlap{\lower 3pt\hbox{$\sim$}}\raise 2.0pt\hbox{$<$}}}
\def\simgt{\mathrel{\rlap{\lower 3pt\hbox{$\sim$}} \raise 2.0pt\hbox{$>$}}}

\def\Msun{M_{\odot}}
\def\Zsun{Z_{\odot}}
 
\newcommand{\q}{\begin{equation}}
\newcommand{\qa}{\begin{eqnarray}}
\newcommand{\qs}{\begin{eqnarray*}}
\newcommand{\nq}{\end{equation}}
\newcommand{\nqa}{\end{eqnarray}}
\newcommand{\nqs}{\end{eqnarray*}}

\makeindex
\begin{document}

\tableofcontents

\clearpage
\newpage

\Chapter[Feedback]{Feedback Processes at Cosmic Dawn}{Ferrara, A. \&  Salvaterra, R.\\
SISSA/ISAS, Via Beirut 4, 34014 Trieste, Italy}

The word ``feedback'' is by far one of the most used ones in modern cosmology where
it is applied to a vast range of situations and astrophysical objects. However, for
the same reason, its meaning in the context is often unclear or fuzzy. Hence a review
on feedback should start from setting the definition of feedback on a solid basis.  
We have found quite useful to this aim to go back to the Oxford Dictionary from where
we take the following definition:

{\it {\bf Fee'dback} n. 1. (Electr.) Return of fraction of output signal from one 
stage of circuit, amplifier, etc. to input of same or preceding stage 
({\bf positive, negative}, tending to increase, decrease the amplification, 
etc). 2. (Biol., Psych., etc) Modification or control of a process or system 
by its results or effects, esp. by difference between the desired and actual 
results.}

In spite of the broad description, we find this definition quite appropriate 
in many ways. First, the definition outlines the fact that the concept of feedback 
invokes a back reaction of a process on itself or on the causes that have produced it.
Secondly, the character of feedback can be either negative or positive. Finally, and most
importantly, the idea of feedback is intimately linked to the possibility that a system
can become self-regulated. Although some types of feedback processes are disruptive, 
the most important ones in astrophysics are probably those that are able to drive the systems towards 
a steady state of some sort. To exemplify, think of a galaxy which is witnessing a burst of star formation. 
The occurrence of the first supernovae will evacuate/heat the gas thus suppressing the star formation
activity. Such feedback is then acting back on the energy source (star formation); it is of a 
negative type, and it could regulate the star formation activity in such a way that only 
a sustainable amount of stars is formed (regulation). However, feedback can fail to produce
such regulation either in small galaxies where the gas can be ejected by the first 
SNe or in cases when the star formation timescale is too short compared to the feedback one.   
As we will see there are at least three types of feedback, and even the stellar feedback 
described in the example above is part of a larger class of feedback phenomena related to the 
energy deposition of massive stars. We then start by briefly describing the key physical ingredients of 
feedback processes in cosmology.
\setcounter{page}{1}

\section{\index{shock!wave} Shock Waves}
A \index{shock} shock represents a ``hydrodynamic surprise'', in the sense that such 
a wave travels faster than signals in the fluid, which are bound to propagate
at the gas sound speed, $c_s$. It is common to define the shock speed, $v_s$, in
terms of $c_s$ by introducing the \index{shock} shock Mach number, $M$:
\q
v_s=Mc_s=M(\gamma P/\rho)^{1/2},
\nq
where $P$ and $\rho$ are the gas pressure and density, respectively.
The change of the fluid velocity $v$ is governed by the momentum equation 
\q
\rho\frac{\partial v}{\partial t}+\rho v\cdot \nabla v=-\nabla P - \frac{1}{8\pi}\nabla B^2 +\frac{1}{4\pi}B\cdot \nabla B,
\nq
where $B$ is the magnetic field. The
density is determined by the continuity equation
\q
\frac{\partial \rho}{\partial t}+v\cdot \nabla \rho=-\rho\nabla v.
\nq
\subsection{Hydrodynamics of \index{shock!wave} Shock Waves}

The large-scale properties of a \index{shock} shock in a perfect gas, with $B=0$ are fully described 
by the three ratios $v_2/v_1$ $\rho_2/\rho_1$ and $P_2/P_1$, determined in 
terms of the conditions ahead of the \index{shock} shock (i.e. $v_1$, $\rho_1$ and $P_1$) by the
Rankine-Hugoniot jump conditions resulting from matter, momentum and energy conservation: 

\q
\rho_1 v_1=\rho_2 v_2,
\nq
\q\label{eq:mom}
P_1 +\rho_1 v_1^2=P_2 +\rho_2 v_2^2,
\nq
\q
v_2 (\frac{1}{2}\rho_2 v_2^2 + U_2)-v_1(\frac{1}{2}\rho_1 v_1^2 + U_1)=v_1 P_1 - v_2 P_2, 
\nq

\noindent
where $U$ is the internal energy density of the fluid. If the fluid behaves
as a perfect gas on each side of the front we have $U=P/(\gamma-1)$. 

In the adiabatic limit it is sufficient to consider the mass and momentum equations only. 
By solving such system we obtain:
\q
\frac{v_2}{v_1}=\frac{\rho_1}{\rho_2}=\frac{\gamma-1}{\gamma+1}+\frac{2}{\gamma+1}\frac{1}{M^2},
\nq
\noindent
where the Mach number is defined as $M^2=v_1^2/c_1^2$ and $c_1$ is the 
sound velocity in the unperturbed medium ahead of the \index{shock} shock. 
For strong \index{shock} shocks, i.e. $M\gg 1$ and $\gamma=5/3$, $\rho_2/\rho_1=4$ and 
\q
T_2=\frac{3\mu}{16 k}v_s^2,
\nq
\noindent
where $v_s=v_1-v_2$

In the isothermal limit, i.e. when the cooling time of the postshock gas
becomes extremely short and the initial temperature is promptly re-established,

\q
\frac{\rho_2}{\rho_1}=\frac{v_1^2}{c_s^2}=M^2,
\nq

\noindent
where $c_s$ is the sound velocity on both sides of the 
\index{shock} shock. While in the adiabatic case the compression factor is limited to four, 
in an isothermal \index{shock} shock much larger compressions are possible due to the softer
equation of state characterizing the fluid. 

\subsection{Hydromagnetic \index{shock!wave} Shock Waves}

Assuming $B$ is parallel to the \index{shock} shock front (so that $B\cdot\nabla B=0$), 
eq. (\ref{eq:mom}) can be replaced by

\q
P_1 +\rho_1 v_1^2 + \left(\frac{B_1^2}{8\pi}\right)=P_2 +\rho_2 v_2^2 + \left(\frac{B_2^2}{8\pi}\right).
\nq

Magnetic flux conservation  requires $B_1/\rho_1=B_2/\rho_2$, so that
in the adiabatic limit the magnetic pressure $B^2/8\pi$ increases by only a
factor 16.

In the isothermal limit the compression depends only linearly on the Alfv\'enic Mach number $M_a$ 

\q
\frac{\rho_2}{\rho_1}=\sqrt{2}\frac{v_1}{v_{a,1}}=M_a,
\nq

\noindent
where $v_a=B/\sqrt{4\pi\rho_1}$ is the Alfv\'en velocity. Hence, some of the \index{shock} shock kinetic energy
is stored in the magnetic field lines rather than being used to compress the gas as in the
field-free case; in other words a magnetized gas is less compressible than an unmagnetized one. 

\subsubsection{Structure of Radiative \index{shock!radiative} Shocks}

The structure of a \index{shock} shock can be divided into four regions 
(McKee \& Draine 1991). The radiative 
precursor is the region upstream of the \index{shock} shock in which radiation emitted by 
the \index{shock} shock acts as a precursor of the \index{shock} shock arrival. The \index{shock} shock front is
the region in which the relative kinetic energy difference of the shocked and unshocked
gas is dissipated. If the dissipation is due to collisions among the atoms or
molecules of the gas, the shock is collisional. On the other hand, if the 
density is sufficiently low that collisions are unimportant and the dissipation
is due to the collective interactions of the particles with the turbulent electromagnetic
fields, the \index{shock} shock is collisionless. Next comes the radiative zone, in which
collisional processes cause the gas to radiate. The gas cools and the density
increases. Finally, if the shock lasts long enough, a thermalization
region is produced, in which radiation from the radiative zone is absorbed and  
re-radiated. 

\subsection{\index{supernova!explosion} Supernova Explosions}

In the explosion of a supernova three different stages in the expansion may
be distinguished. In the initial phase the interstellar material has little 
effect because of its low moment of inertia; the velocity of expansion of the 
supernova envelope will then remain nearly constant with time. This phase terminates
when the mass of the swept up gas is about equal to the initial mass $M_e$ 
expelled by the supernova, i.e. $(4\pi/3)r_s^3\rho_1=M_e$, where $\rho_1$
is the density of the gas in front of the \index{shock} shock and $r_s$ is the radius
of the shock front. For $M_e=0.25\;\Msun$ and $\rho_1=2\times10^{-24}$ 
g/cm$^3$, $r_s\simeq1$ pc, which will occurs about 60 yr after the explosion.
During the second phase (Sedov-Taylor phase) the mass behind the \index{shock} shock is 
determined primarily by the amount of interstellar gas swept up, but the
energy of this gas will remain constant. When radiative cooling becomes important  
(radiative phase), the temperature of the gas will fall to a relatively low
value. The motion of the \index{shock} shock is supported by the momentum of the
outward moving gas and the \index{shock!radiative} shock may be regarded as isothermal. The velocity 
of the shell can be computed from the condition of momentum conservation 
(snowplow model). This phase continues until $v_s=\max\{v_t,c_s\}$, where $v_t$ is
the turbulent velocity of the gas, when the shell loses its identity due to gas random 
motions.

\subsubsection{Blastwave Evolution}
To describe the evolution of a blastwave, the so-called thin shell approximation is frequently
used. In this approximation, one assumes that most of the mass of the material swept up 
during the expansion is collected in a thin shell. We start by applying
(we neglect the ambient medium density and shell self-gravity) the virial theorem to the shell:

\q\label{eq:VT}
\frac{1}{2}\frac{d^2 I}{dt^2}=2E=2(E_k + E_t),
\nq
where $I$ is the moment of inertia, and $E_k$ ($E_t$) is the kinetic (thermal)
energy of the shell.
Let us introduce the structure parameter $K$ in the following form:

\q
K=\frac{1}{MRv_2}\int_0^R dm\,rv,
\nq

\noindent
where $M$ is the ejected mass, $R$ the shell radius. Substituting in 
equation~(\ref{eq:VT}) with $v=v_2(r/R)$, we obtain

\q
K \frac{d}{dt}(Mrv_s)=2E(v_s/v_2).
\nq

To recover the Sedov-Taylor adiabatic phase we integrate up to $R$ and assume
$\rho(R)=\rho_0 R^{-m}$, $E=const.$, $K=(6-2m)/(7-2m)$

\q
R(t)=\left[\frac{(5-m)E}{\pi \rho_0 K}\right]^{1/(5-m)} t^{2/(5-m)}.
\nq
The previous expression can be obtained also via a dimensional approach. In fact, in the adiabatic phase
the energy is conserved and hence 
\q
E\propto Mv_s^2\propto \rho_0 R^{5-m}t^{-2},
\nq
so that

\q
R^{5-m}\propto \frac{E}{\rho_0}t^2;
\nq

\noindent
the fraction of the total energy transformed in kinetic form is

\q
E_k=\frac{1}{2}K M \left(\frac{3}{4}v_s^2\right)=\frac{3E}{2(5-m)}.
\nq
We note that in the limit of a homogeneous ambient medium ($m=0$) we recover
the standard Sedov-Taylor evolution $R \propto t^{2/5}$, with 30\% of the
explosion energy in kinetic form.

A similar reasoning can be followed to recover the radiative phase behavior. 
In this phase $v_2=v_s$, i.e. the \index{shock!radiative} shock
wave is almost isothermal and $K=1$, i.e. the shell is thin.
Under these assumptions the energy is lost at a rate

\q
\dot{E}=-4\pi R^2 \left(\frac{1}{2}\rho_0 v_s^2\right).
\nq

There are two possible solutions of equation~(\ref{eq:VT}) in this case:
(a) the pressure-driven snowplow in which

\q
R(t)\propto t^{2/(7-m)},
\nq

\noindent
or the momentum-conserving solution
\q
R(t)\propto t^{1/(4-m)}.
\nq
In both cases, the expansion proceeds at a lower rate with respect to
the adiabatic phase.

\begin{figure}
\begin{center}
\includegraphics[width=12cm]{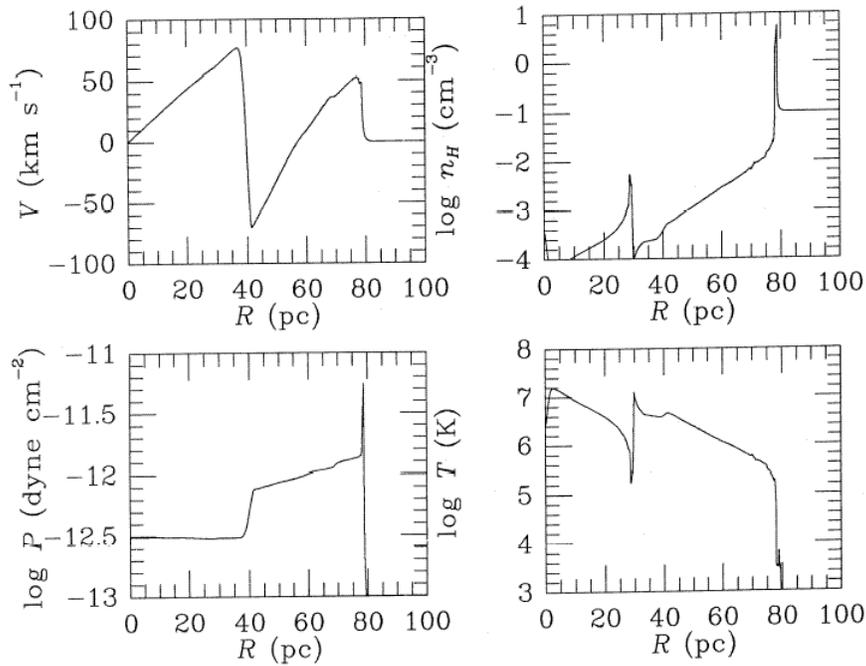}
\end{center}
\caption{The structure of spherical supernova remnant calculated using a
high-resolution numerical hydrodynamical code. The remnant age is 
$t=5\times 10^5$~yr. This Figure is taken from Cioffi, McKee \& Bertschiger 
(1988)}
\label{fig:remnant}
\end{figure}

In order to study the evolution of an explosion in a more realistic way
it is necessary to resort to numerical simulations as the one shown in 
Figure~\ref{fig:remnant}. From there the inner structure of the remnant is 
clearly visible:
indeed most of the mass is concentrated in a thin shell (located at about 80 pc
after 0.5 Myr from the explosion); a reverse shock is also seen which is proceeding
towards the center and thermalizing the ejecta up to temperatures of the order of $10^7$~K. 
In between the two \index{shock} shocks is the
contact discontinuity between the ejecta and the ambient medium, through which the
pressure remains approximately constant.   

\section{\index{ionization!photo-} Photoionization}

The ionization balance equation around a source is given by 

\q
4\pi R^2 n \frac{dR}{dt}+\frac{4}{3}\pi R^3 n^2 \alpha = 4\pi S_i(0),
\nq

\noindent
where $S_i(0)$ is the \index{ionizing photon} ionizing photon rate from the source, $n$ is the 
neutral gas density and $\alpha$ is the recombination coefficient. 
\noindent
The solution of the ionization balance equation is

\q
R(t)=R_S[1-\exp(-2n\alpha t/3)]^{1/3}.
\nq

\noindent
As time goes to infinity the radius tends to the \index{Stromgren radius} Stromgren radius

\q
R_S= \left(\frac{3S_i(0)}{\alpha n^2}\right)^{1/3}.
\nq

\noindent
The above equations implies that $R_S$ is reached in 

\q
t_S=\frac{3}{2}n\alpha\simeq 2\times 10^5 n \mbox{ yr},
\nq

\noindent
and the typical \index{ionization!front} ionization front (I-front) velocity is $v_I\propto S_i(0)/R^2n$.

Depending on the value of $v_I$ we can distinguish three types of I-fronts,
where $c_i$ ($c_n$) is the sound speed in the ionized (neutral) gas:

\begin{enumerate}
\item R(arefied)-TYPE: $v_I>(c_i,c_n)$
\item Strong D(ense)-TYPE: $c_n< v_I<c_i$
\item Weak D(ense)-TYPE: $v_I< (c_i,c_n)$
\end{enumerate}

In the case of R-TYPE I-front there is no hydrodynamical reaction to the 
passage of the ionizing front, since the its motion is supersonic relative
to the gas behind as well as ahead of the front. In the case of D-TYPE fronts
the dynamical timescale becomes short enough for the gas to react to \index{ionization!photo-} photoionization
and a \index{shock} shock will form and precede the \index{ionization!front} ionizing front thus increasing the ambient gas density.

\subsubsection{\index{ionization!photo-} Photoionization Equilibrium}

It is convenient to divide the ionizing flux into two parts, a `stellar'
part $S(\nu)$, resulting directly from the input radiation from the star,
and a `diffuse' part $S_d(\nu)$, resulting from the emission of the 
ionized gas, that in a pure H gas is due to recombinations into 1$^2$ S
state. 

\q
S_t(\nu)=S(\nu)+S_d(\nu).
\nq

\noindent
Both the recombination radiation and photoionization cross section 
($\sigma\propto \nu^3$) are peaked at the threshold, $\nu_L$. We can
identify two approximate cases (Osterbrok 1989):

\begin{enumerate}
\item MENZEL CASE A: optically thin nebula, i.e. 
$\tau(\nu=\nu_L)=n_H\sigma(\nu_L)R<1$, and the diffuse photons can escape;
\item MENZEL CASE B: optically thick nebula, i.e. diffuse photons are 
absorbed very close to the point at which they are generated (`on-the-spot' 
approximation).
\end{enumerate}

Menzel case B is usually a good approximation for many astrophysical purposes
because the diffuse radiation-field photons have $\nu\simeq \nu_L$, and therefore 
a short mean free path before \index{absorption} absorption. Under the hypothesis of the on-the-spot 
approximation, the recombination coefficient is given by

\q
\alpha_{eff}=\alpha-\alpha_{1S},
\nq

\noindent
resulting roughly in a 60\% decrease. The physical meaning is that in optically thick 
nebulae, the ionizations caused by stellar radiation-field photons are 
balanced by recombinations to excited levels of H, while recombinations to the
ground level generate \index{ionizing photon} ionizing photons that are absorbed elsewhere in the 
nebula but have no net effect on the overall ionization balance.

\section{\index{thermal instability} Thermal Instability}\label{sec:theq}

A final crucial ingredient to understand feedback and self-regulation is 
represented by the thermal instability. 
Let us rewrite the usual mass, momentum and energy equations in a slightly
different form (Balbus 1995):

\q
\frac{d\ln\rho}{dt}+\rho \nabla \cdot v=0,
\nq
\q
\rho \frac{dv}{dt}=-\nabla P,
\nq
\q
\frac{3}{2}P\frac{d\ln P}{dt}\rho^{-5/3}=-\rho L(\rho,T),
\nq

\noindent
where $L$ is the net \index{cooling!function} cooling function (i.e. heating-cooling)  and is taken to be a function of density and
temperature; in the equilibrium state $v=L=0$. We perturb the equilibrium 
by introducing small disturbance. We write such perturbations as 
$\delta q\propto \mbox{e}^{(kr+nt)}$; $nt$ is then the growth rate of 
the instability. The linearized system yields the dispersion relation 

\q\label{eq:linear}
n^3+\frac{2}{3}T\Theta_{T,\rho}n^2+k^2c_s^2n+\frac{2}{5}k^2c_s^2T\Theta_{T,\rho}=0,
\nq

\noindent
where $\Theta\equiv (\rho/P)L$ and 
$\Theta_{A,B}\equiv (\partial \Theta/\partial A)_B$.

For large $k$ (small wavelengths) the instability occurs isobarically, that is
pressure is kept approximately constant as the density increases following
the process of condensation and cooling. In this case such equations has three solutions:

\q
n_1=-\frac{2}{5}T \Theta_{T,P},
\nq
\q
n_{2,3}=\pm ikc_s-\frac{2}{15}T \Theta_{T,S}.
\nq

\noindent
It is straightforward to recognize that $n_{2,3}$ are the modified sound wave modes,
whereas $n_1$ corresponds to the condensation mode.
Note also that the sign of the temperature derivative of the \index{cooling!function} cooling function
determines whether or not a mode is unstable (instability occurs if $\Theta_{T,P}<0$).

In the opposite (isochoric) limit characterized by small $k$ (large wavelengths), 
the condensation mode has the following expression:

\q
n_{1}=-\frac{2}{3}T\Theta_{T,\rho},
\nq

\noindent
Although formally a solution of the dispersion equation, the \index{thermal instability} thermal instability
cannot physically occur under isochoric conditions. The reason is that 
on such large scales there is not enough time to establish pressure equilibrium as
the cooling time is much shorter than the sound crossing time over a wavelength. 
Thus the density remains approximately constant while the gas cools, 
and pressure equilibrium is likely to be restored promptly and rather dramatically  
by a propagating \index{shock} shock front.

\begin{figure}
\begin{center}
\includegraphics[width=12cm]{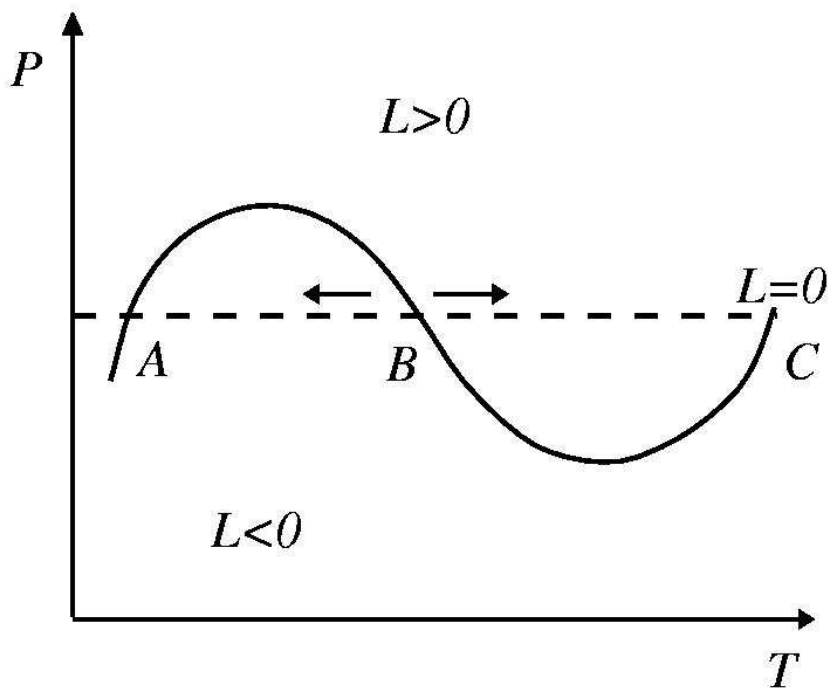}
\end{center}
\caption{Equilibrium solution curve $L=0$ in the $P-T$ plane for a hypothetical
cooling function. See text.}
\label{fig:P-T}
\end{figure}

To better understand the nature of the \index{thermal instability} thermal instability let us 
plot the curve of the equilibrium solution $L=0$ in the $P-T$ plane 
for a hypothetical \index{cooling!function} cooling function (Figure~\ref{fig:P-T}). Above the curve,
heating dominates over cooling; below $L=0$, the gas is cooled. 
Between $P_{min}$ and $P_{max}$
a constant pressure line  will intersect the curve in three points (A, B, and C); A and B
are thermally stable. This is readily understood since a small isobaric 
increase (decrease) in temperature at either of these points takes the system
into the cooling (heating) region of the $P-T$ plane from where it is immediately 
driven back to equilibrium position. On the other hand, a small increase (decrease) in the
temperature in the unstable point B takes one into the heating (cooling) 
region, with the result that the system is driven yet further from its 
equilibrium. This leads to the coexistence of two (or more) thermodynamic
phases and provides the main argument to interpret the multiphase medium commonly
observed in galaxies and in cosmic gas. We will see later on how this concept is 
closely linked with the self-regulation connected to feedback processes.

\section{The Need for Feedback in Cosmology}

One of the main aims of physical cosmology is to understand in detail
galaxy formation starting from the initial density fluctuation field
with a given spectrum (typically a CDM one). Such {\it ab initio} computations
require a tremendous amount of physical processes to be included before it
becomes possible to compare their predictions with experimental data. In particular,
it is crucial to model the interstellar medium of galaxies, which is know to be
turbulent, have a multi-phase thermal structure, can undergo gravitational instability 
and form stars. To account for all this complexity, in addition one should treat 
correctly all relevant cooling processes, radiative and \index{shock} shock heating (let alone 
magnetic fields!). This has proven to be essentially impossible even with present day
best supercomputers. Hence one has to resort to heuristic models where simplistic 
prescriptions for some of these processes must be adopted. Of course such an approach
suffers from the fact that a large number of free parameters remains which cannot be
fixed from first principles. These are essentially contained in each of the ingredients
used to model galaxy formation, that is, the evolution of dark halos, cooling and star
formation, chemical enrichment, stellar populations. Fortunately, there is a large
variety of data against which the models can be tested: these data range from the
fraction of cooled baryons to cosmic star formation histories, the luminous content
of halos, luminosity functions and faint galaxy counts. The feedback processes enter the 
game as part of such iterative try-and-learn process to which we are bound by our
ignorance in dealing with complex systems as the galaxies. Still, we are far from 
a full understanding of galaxies in the framework of structure formation models.      
The hope is that feedback can help us to solve some ``chronic'' problems found in
cosmological simulations adopting the CDM paradigm.  Their (partial) list includes: 

\begin{enumerate}
\item {\bf Overcooling}: the predicted cosmic fraction of cooled baryons is larger
than observed. Moreover models predict too many faint, low mass galaxies;
\item {\bf Disk Angular Momentum}: the angular momentum loss is too high and 
galactic disk scale lengths are too small;
\item {\bf Halo Density Profiles}: profiles are centrally too concentrated;
\item {\bf Dark Satellites}: too many satellites predicted around our Galaxy.
\end{enumerate}


\subsection{The Overcooling Problem}

Among the various CDM problems, historically the overcooling has been the most
prominent and yet unsolved one. In its original formulation it has been first spelled out by White \& Frenk (1991).
Let us assume that, as a halo forms, the gas initially relaxes to an isothermal
distribution which exactly parallels that of the dark matter. The virial
theorem then relates the gas temperature $T_{vir}$ to the 
circular velocity of the halo $V_c$,

\q
kT_{vir}=\frac{1}{2}\mu m_p V_c^2 \mbox{ or } T_{vir}=36 V^2_{c,km/s} \mbox{K},
\nq

\noindent
where $\mu m_p$ is the mean molecular weight of the gas. At each radius in 
this distribution we can then define a cooling time as the ratio of the 
specific energy content to the cooling rate,

\q
t_{cool}(r)=\frac{3\rho_g(r)/2\mu m_p}{n_e^2(r)\Lambda(T)},
\nq

\noindent 
where $\rho_g(r)$ is the gas density profile and $n_e(r)$ is the electron 
density. $\Lambda(T)$ is the \index{cooling!function} cooling function. The \index{cooling!radius} cooling radius is defined
as the point where the cooling time is equal to the age of the universe, i.e.
$t_{cool}(r_{cool})=t_{Hubble}=H(z)^{-1}$.

Considering the virialized part of the halo to be the region encompassing a
mean overdensity is 200, its radius and mass are defined by

\q
r_{vir}=0.1H_0^{-1}(1+z)^{3/2}V_c,
\nq
\q\label{eq:mvir}
M_{vir}=0.1(GH_0)^{-1}(1+z)^{3/2}V_c^3.
\nq

Let us distinguish two limiting cases. When $r_{cool}\gg r_{vir}$ (accretion
limited case), cooling is so rapid that the infalling gas never comes to 
hydrostatic equilibrium. The supply of cold gas for star formation is then
limited by the infall rate rather than by cooling. The accretion rate is obtained
by differentiating equation~(\ref{eq:mvir}) with respect to time and multiplying
by the fraction of the mass of the universe that remains in gaseous form:

\q\label{eq:macc}
\dot{M}_{acc}=f_g\Omega_g \frac{d}{dt} 0.1 (GH_0)^{-1}(1+z)^{3/2}V_c^3= 
0.15 f_g \Omega_g G^{-1}V_c^3.
\nq

\noindent 
Note that, except for a weak time dependence of the fraction
of the initial baryon density which remains in gaseous form, $f_g$, this
infall rate does not depend on redshift. 

In the opposite limit, $r_{cool}\ll r_{vir}$ (quasi-static case), the 
accretion \index{shock} shock radiates only weakly, a quasi-static atmosphere forms, and
the supply of cold gas for star formation is regulated by radiative losses
near $r_{cool}$. A simple expression for the inflow rate is 

\q\label{eq:mqst}
\dot{M}_{qst}=4\pi\rho_g(r_{cool})r_{cool}^2 \frac{d}{dt}r_{cool}.
\nq

\begin{figure}
\begin{center}
\includegraphics[width=12cm]{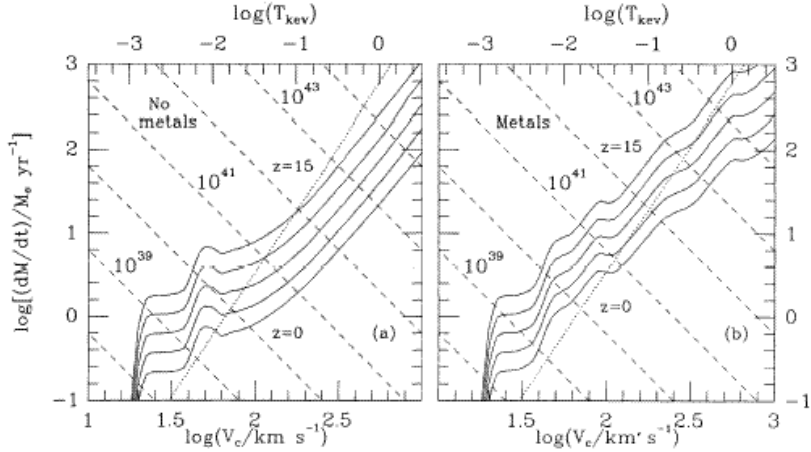}
\end{center}
\caption{Gas infall rate and cooling rates in dark
matter halos as a function of circular velocity and redshift. The infall
rate ({\it dotted line}) is essentially independent of redshift; the cooling
rates ({\it solid lines}) are given for redshift $z=0,1,3,7$ and 15 ({\it from
bottom to top}). Dashed lines give present-day X-ray luminosities in erg 
s$^{-1}$ produced by gas cooling at the given rate in each halo. The 
predicted temperature of this emission is given on the upper abscissa. 
{\it (a)} A \index{cooling!function} cooling function for gas of zero metallicity is assumed. 
{\it (b)} A cooling function for metal enriched gas. 
This Figure is taken from White \& Frenk (1991).} 
\label{fig:rates}
\end{figure}

The gas supply rates predicted by equations~(\ref{eq:macc}) and 
(\ref{eq:mqst}) are illustrated in Figure~\ref{fig:rates}.
In any particular halo, the rate at which cold gas becomes available for star
formation is  the minimum between $\dot{M}_{acc}$ and $\dot{M}_{qst}$.

According to the models of Thomas et al. (1986), the bolometric X-ray 
luminosity of the region within the \index{cooling!radius} cooling radius of a galactic cooling flow
is

$$
L_X=2.5 \dot{M}_{cool}V_c^2.
$$

The predictions of this formula are superposed on Figure~\ref{fig:rates} 
(dashed lines). For large circular velocities the mass cooling rates 
correspond to quite substantial X-ray luminosities. 

Integrating the gas supply rates $\dot{M}_{cool}$ over redshift and 
halo mass distribution, we find that for $\Omega_b=0.04$ {\it most of the gas
is used before the present time}. This is unacceptable, since the density contributed
by the observed stars in galaxies is less than 1\% of the critical density. 
So, the star formation results to be too rapid without a regulating process,
i.e. feedback. 

To solve this puzzle, Larson (1974) proposed that the energy input from young stars
and supernovae could quench star formation in small protogalaxies before more
than a small gas fraction has been converted into stars. 

Stellar energy input would counteract radiative losses in the cooling gas and
tend to reduce the supply of gas for further star formation. 
We can imagine that the star formation process is self-regulating in the sense
that the star formation rate $\dot{M}_\star$ takes the value required for heating to balance 
dissipation in the material which does not form stars. This produces the 
following prescription for the star formation rate:

\qa
\dot{M}_\star(V_c,z)& =& \epsilon(V_c) \mbox{min}(\dot{M}_{acc},\dot{M}_{qst}),\\
\epsilon(V_c)& =& [1+\epsilon_0(V_0/V_c)^2]^{-1}.
\nqa

\noindent
For large $V_c$ the available gas turns into stars with high efficiency because
the energy input is not  sufficient to prevent cooling and fragmentation; for
smaller objects the star formation efficiency $\epsilon$ is proportional to 
$V_c^2$. The assumption of self regulation at small $V_c$ seems plausible 
because the time interval between star formation and energy injection is much
shorter than either the sound crossing time or the cooling time in the 
gaseous halos. However, other possibilities can be envisaged. For example, 
Dekel \& Silk (1986) suggested that supernovae not only would suppress cooling
in the halo gas but would actually expel it altogether. The conditions leading 
to such an event are discussed next. 

\subsection{\index{dwarf galaxy} Dwarf Galaxies: a Feedback Lab}

The observation of dwarf galaxies has shown well-defined correlations 
between their measured properties, and in particular 

\begin{enumerate}
\item Luminosity-Radius $L\propto R^r$ with $r=4$
\item Luminosity-Metallicity $L\propto Z^z$ with $z=5/2$
\item Luminosity-Velocity dispersion $L\propto V^v$ with $v=4$
\end{enumerate}

A simple model can relate the observed scaling parameters $r$, $z$, and $v$
to each other  (Dekel \& Silk 1986).
Consider a uniform cloud of initial mass 
$M_i=M_g+M_\star$ ($M_g$ is the mass of gas driven out of the system, and 
$M_\star$ is the mass in stars) in a sphere of radius $R_i$ which undergoes
star formation. The metallicity for a constant yield in the instantaneous
recycling approximation is given by $Z=y\ln(1+M_\star/M_g)$ where $y$ are
the yields.

Let us impose simple scaling relations: 
$M_i\propto M$ (gas mass proportional to the dark matter mass), 
$R\simeq R_i$ , and $V\simeq V_i$ (gas loss has no dynamical effect).
From the observed scaling relation we have 
$L\propto R^r\propto V^v$. Now we should write the analogous relations 
for structure and velocity that hold before the removal, i.e. 
$M_i\propto R^{ri}$ and $M_i\propto V^{vi}$.

Consider now the case in which the gas is embedded in a dark halo, and
assume that when it forms stars the mass in gas is proportional to the
dark matter inside $R_i$, $M_i\propto M$. If the halo is dominant, the gas
loss would have no dynamical effect on the stellar system that is left 
behind, so $R\simeq R_i$ and $V\simeq V_i$. The relations for structure and 
velocity that hold before the removal give

\q
L\propto R^r\propto V^v, \;\; M\propto R^{ri}, \;\; V^2\propto M/R,
\nq

\noindent
so that we obtain

\q
v=2r/(ri-1).
\nq

In the limit $M_\star\ll M_g$ and $L\propto M_\star$, from the metallicity
relation we have

\q
Z\propto \frac{M_\star}{M_g}\propto \frac{L}{M}\propto L^{1/z},
\nq

\noindent
so that

\q
r/ri=(z-1)/z.
\nq

The final equation comes from the energy condition. For thermal energy, in 
the limit of substantial gas loss, $M_g\simeq M_i$, we have $L\propto MV^2$, 
and so

\q
v=2z.
\nq

For a CDM spectrum $P(k)=Ck^{n_s}$, we obtain $r_i=6/(5+n_s)$
and introducing $z=2r/(r-1)$ and $n_s=12/(r+1)-5$. For $r=4$ we find $n_s=-2.6$
consistent with the CDM and the scaling relations are

\q
L\propto R^4\propto Z^{2.7}\propto V^{5.3}.
\nq

Let us now investigate the critical conditions, in terms of gas density $n$
and virial velocity $V$, for a global supernova-driven gas removal from a 
galaxy while it is forming stars. Here, spherical symmetry and the presence of
a central point source are assumed.  The basic requirement for gas removal 
is that the energy that has been pumped into the gas is enough to expel it
from the protogalaxy. The energy input in turn depends on the supernova rate,
on the efficiency of energy transfer into the gas, and on the time it takes for the SNRs 
to overlap and hence affect a substantial fraction of the gas. The first
is determined by the rate of star formation, the second by the evolution 
of the individual SNRs, and the third by both.
When all these are expressed as a function of $n$ and $V$, the critical condition
for removal takes the form

\q
E(n,V)\geq \frac{1}{2}M_g V^2.
\nq

\noindent
This relation defines a locus in the $n-V$ diagram shown in 
Figure~\ref{fig:removal} within which substantial gas loss is possible. 
In Figure~\ref{fig:removal}, the cooling
curve, above which the cooling time is less than the free-fall time, confines
the region where the gas can contract and form stars. The almost vertical
line $V_{crit}$ divides the permissible region for galaxy formation in two;
a protogalaxy with $V>V_{crit}$ would not expel a large fraction of its
original gas but rather turn most of its original gas into stars to form a
`normal' galaxy. A protogalaxy with $V<V_{crit}$ can produce a 
supernova-driven wind out of the first burst of star formation, which would
drive a substantial fraction of the protogalactic gas out, leaving a diffuse
\index{dwarf galaxy} dwarf. 

The short-dashed curve marked ``1$\sigma$'' corresponds to density
perturbations $\delta M/M$ at their equilibrium configuration after a 
dissipationless collapse from a CDM spectrum, normalized to $\delta M/M$ at 
a comoving radius $8h^{-1}$ Mpc. The density $n$ is calculated for a uniformly
distributed gas in the CDM halos, with a gas-to-total mass ratio $\chi=0.1$.
The corresponding parallel short-dashed curve corresponds to the protogalactic
gas clouds, after a contraction by a factor $\chi^{-1}=10$ inside isothermal
halos, to densities such that star
formation is possible. The vertical dashed arrow marks the largest galaxy
that can form out of a typical 1$\sigma$ peak in the initial distribution
of density fluctuations. The vast majority of such protogalaxies, when they
form stars, have $V<V_{crit}$, so they would turn into \index{dwarf galaxy} dwarfs. 
The locus where ``normal'' galaxies are expected to be found is the shaded
area. It is evident that most of them must originate form 2$\sigma$ and
3$\sigma$ peaks in the CDM perturbations. 

So, the theory hence predicts two distinct types of galaxies which occupy
two distinct loci in the $n-V$ diagram: the ``normal'' galaxies are confined
to the region of larger virial velocities and higher densities, and they tend
to be massive; while the diffuse dwarfs are typically of smaller velocities
and lower densities, and their mass in star is less than $5\times10^9\;\Msun$.

\begin{figure}
\begin{center}
\includegraphics[width=12cm]{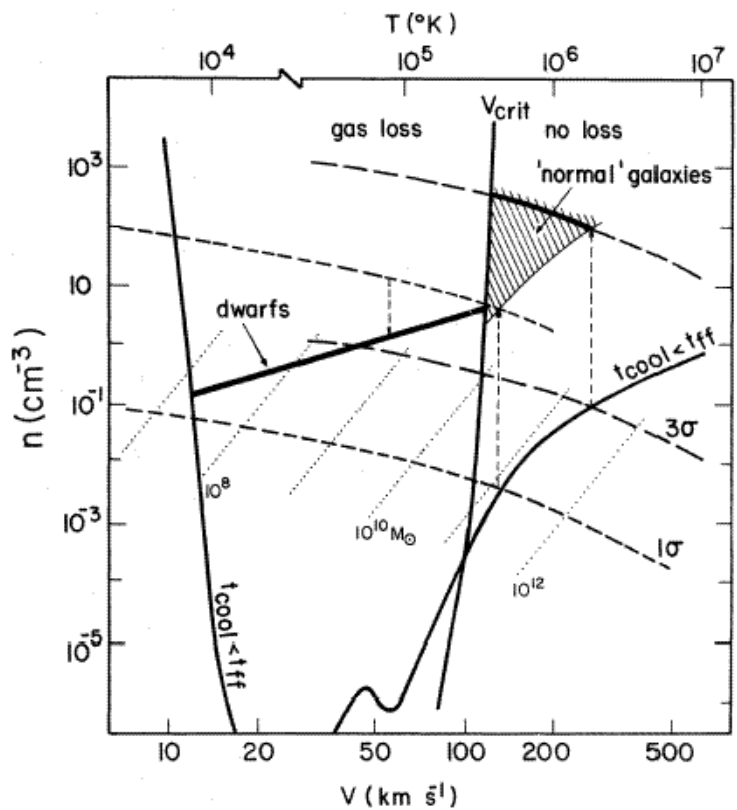}
\end{center}
\caption{Gas number density vs. virial
velocity (or viral temperature), the formation of dwarf vs. `normal' galaxies
in CDM halos, and the origin of biased galaxy formation. This Figure is 
taken from Dekel \& Silk (1986).}
\label{fig:removal}
\end{figure}

This simple model, in spite of its enlightening power, has clear limitations.
The most severe is that is assumes a spherical geometry and a single \index{supernova!explosion} supernova
explosion. These assumptions have been released by a subsequent study based on 
a large set of numerical simulations (Mac-Low \& Ferrara 1999).
These authors modelled the effects of repeated supernova explosions from starbursts
in \index{dwarf galaxy} dwarf galaxies on the interstellar medium of these
galaxies, taking into account the gravitational potential of their
dominant dark matter halos.  They explored supernova rates from one every
30,000 yr to one every 3 Myr, equivalent to steady mechanical
luminosities of $L=0.1-10 \times 10^{38}$~ergs~s$^{-1}$, occurring in
dwarf galaxies with gas masses $M_g=10^6 - 10^9 M_\odot$. Surprisingly, Mac-Low
\& Ferrara (1999)
found that the mass ejection efficiency is very low for galaxies with
mass $M_g \geq 10^7 M_\odot$.  Only galaxies with $M_g \simlt 10^6
M_\odot$ have their interstellar gas blown away, and then virtually
independently of $L$ (see Table \ref{tab:mass}).  
On the other hand, metals from the supernova
ejecta are accelerated to velocities larger than the escape speed from
the galaxy far more easily than the gas. They found\footnote{Throughout the text we use the 
standard notation $Y_X = Y/10^X$ } that for $L_{38}=1$, only
about 30\% of the metals are retained by a $10^9 M_\odot$ galaxy, and
virtually none by smaller galaxies (see Table \ref{tab:metal}).

\begin{figure}
\begin{center}
\includegraphics[width=12cm]{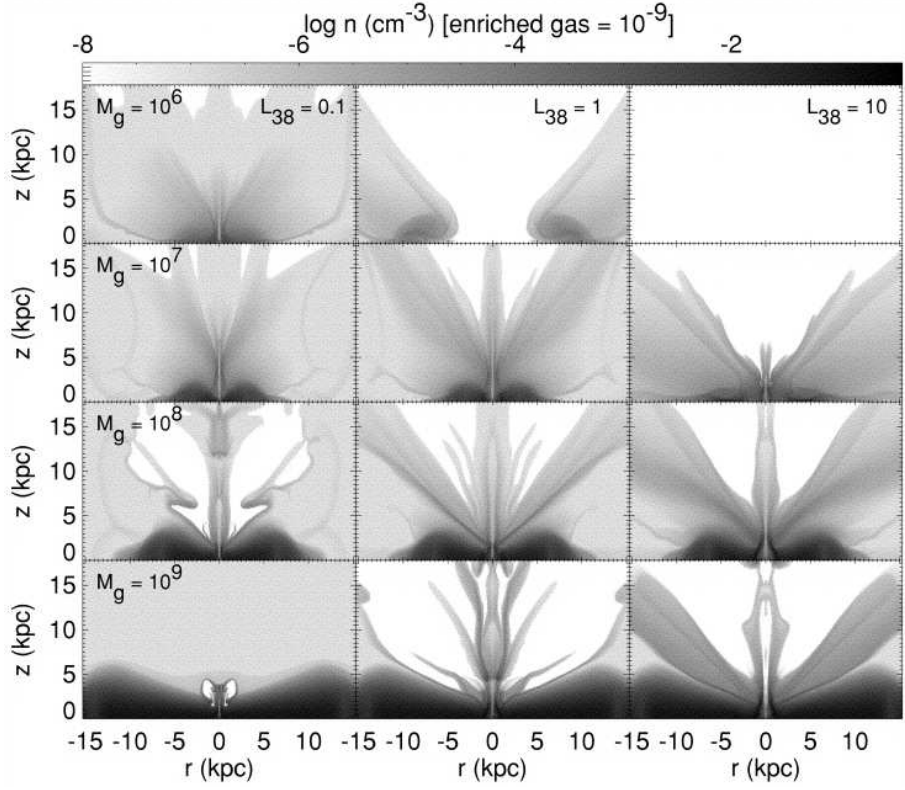}
\end{center}
\caption{Distribution of metal-enriched stellar outflow material and SN ejecta from the starburst energy source at time 200 Myr. Note that in most cases, no enriched gas remains in the disks of the galaxies. This Figure is 
taken from Mac-Low \& Ferrara (1999).}
\label{fig:removal}
\end{figure}

\begin{table}
\begin{center}
\begin{tabular}{crrr}
\br
Visible Mass & \multicolumn{3}{c}{Mechanical Luminosity [$10^{38}$ erg s$^{-1}$]} \\
\cline{2-4}
[$\Msun$] & 0.1 & 1.0 & 10 \\
\mr 
 $10^6$ & 0.18 & 1.0 & 1.0\\
 $10^7$ & $3.5\times 10^{-3}$ & $8.4\times 10^{-3}$ & $4.8\times 10^{-2}$ \\
 $10^8$ & $1.1\times 10^{-4}$ & $3.4\times 10^{-4}$ & $1.3\times 10^{-3}$ \\
 $10^9$ & $0.0$ & $7.6\times 10^{-6}$ & $1.9\times 10^{-5}$ \\
\br
\end{tabular}
\end{center}
\caption{Mass ejection efficiency}
\label{tab:mass}
\end{table}

\begin{table}
\begin{center}
\begin{tabular}{crrr}
\br
Visible Mass & \multicolumn{3}{c}{Mechanical Luminosity [$10^{38}$ erg s$^{-1}$]} \\
\cline{2-4}
[$\Msun$] & 0.1 & \0\0\0\0\0\0\0\0\0 1.0 & 10 \\
\mr 
$10^6$ & 1.0 & $1.0$ & 1.0 \\
$10^7$ & 1.0 & $1.0$ & 1.0 \\
$10^8$ & 0.8 & $1.0$ & 1.0 \\
$10^9$ & 0.0 & $0.69$ & 0.97 \\
\br
\end{tabular}
\end{center}
\caption{Metal ejection efficiency}
\label{tab:metal}
\end{table}

\subsection{\index{blowout} Blowout, \index{blowaway} Blowaway and Galactic Fountains}

\begin{figure}
\begin{center}
\includegraphics[width=12cm]{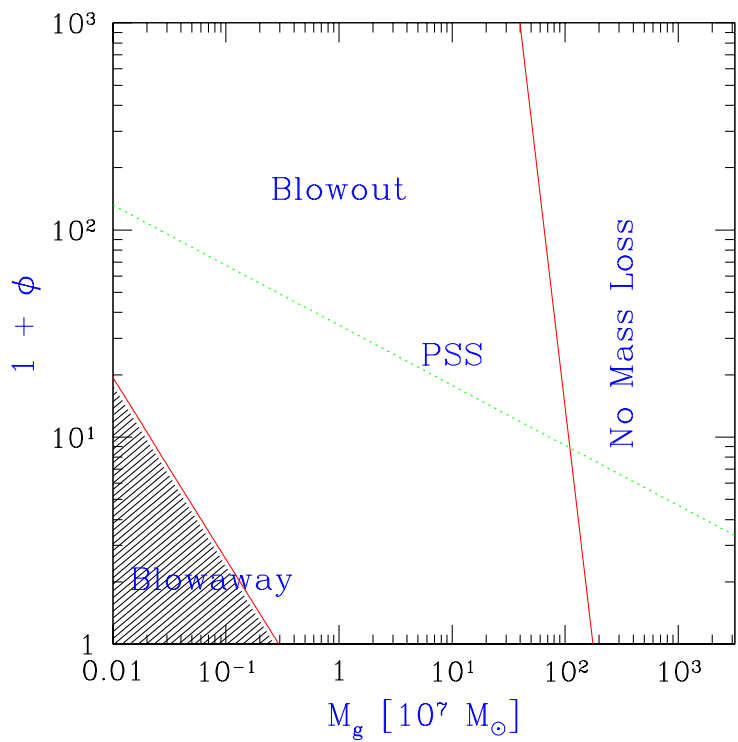}
\end{center}
\caption{Regions in the 
$(1+\phi)-M_{g,7}$ plane in which different dynamical phenomena may occur.
Also shown is the locus point describing the Persic et al. (PSS) relation
(dotted line). This Figure is taken from Ferrara \& Tolstoy (2000).}
\label{fig:eline3}
\end{figure}

The results of the MacLow \& Ferrara study served to clearly classify the various
events induced by starburst in galaxies. 
We can distinguish between blowout and blowaway processes, depending on the 
fraction of the parent galaxy mass involved in the mass loss phenomena.
Whereas in the \index{blowout} blowout process the fraction of mass involved is the one 
contained in cavities created by the \index{supernova!explosion} supernova or 
\index{superbubble} superbubble explosions, the
\index{blowaway} blowaway is much more destructive, resulting in the complete expulsion of the
gas content of the galaxy. 
The two processes lie in different regions of the $(1+\phi)-M_{g,7}$ plane 
shown in figure~\ref{fig:eline3}, where $\phi=M_h/M_g$ is the dark-to-visible 
mass ratio and $M_{g,7}=M_g/10^7\;\Msun$ (Ferrara \& Tolstoy 2000).
Galaxies with gas mass content larger than $10^9\;\Msun$ do not
suffer mass losses, due to their large gravitational well. Of course, this 
does not rule out the  possible presence of outflows with velocities below
the escape velocity (fountain) in which material is temporarily stored in 
the halo and then returns to the main body of the galaxy. For galaxies with
gas mass lower than this value, outflows cannot be prevented. If the mass is
reduced further, and for $\phi\simlt 20$, a \index{blowaway} blowavay, and therefore a complete
\index{gas stripping} stripping of the galactic gas, should occur. To exemplify, the expected 
value of $\phi$ as function of $M_g$ empirically derived by Persic et al. (1996)
is also plotted, which should give an idea of a likely location of 
the various galaxies in the $(1+\phi)-M_{g,7}$ plane.

\subsubsection{\index{blowout} Blowout}

The evolution of a point explosion in an exponentially stratified medium 
can be obtained from dimensional analysis. Suppose that the gas density 
distribution is horizontally homogeneous and that $\rho(z)=\rho_0\exp(-z/H)$.
The velocity of the \index{shock} shock wave is $v\sim(P/\rho)^{1/2}$, where the pressure
$P$ is roughly equal to $E/z^3$, and $E$ is the total energy if the 
explosion. Then it follows that

\q
v(z)\simeq E^{1/2}\rho_0^{-1/2}\exp(z/2H)z^{-3/2}.
\nq

\noindent
This curve has a minimum at $z=3H$ and this value defines the height at which 
the \index{shock} shock wave, initially decelerating, is accelerated to infinity and a 
\index{blowout} blowout takes place. Therefore, $3H$ can be used as the fiducial height where 
the velocity $v_b=v(3H)$ is evaluated. 

There are three different possible fates for SN-shocked gas, depending on the
value of $v_b$. If $v_b<c_{s,eff}$, where $c_s$ is the sound speed in the \index{ISM} ISM,
then the explosion will be confined in the disk and no mass-loss will occur;
for $c_{s,eff}<v_b<v_e$ ($v_e$ is the escape velocity) the supershell will 
breakout of the disk into the halo, but the flow will remain bound to the 
galaxy; finally, $v_e<v_b$ will lead to a true mass-loss from the galaxy.

\subsubsection{\index{blowaway} Blowaway}

The requirement for blowaway is that the momentum of the shell
is larger than the momentum necessary to accelerate the gas outside the shell
at velocity larger than the escape velocity, $M_sv_c \leq M_0v_e$. 
Defining the disk axis ratio as $\epsilon=R/H\geq 1$, the blowaway condition can
be rewritten as

\begin{figure}
\begin{center}
\includegraphics[width=12cm]{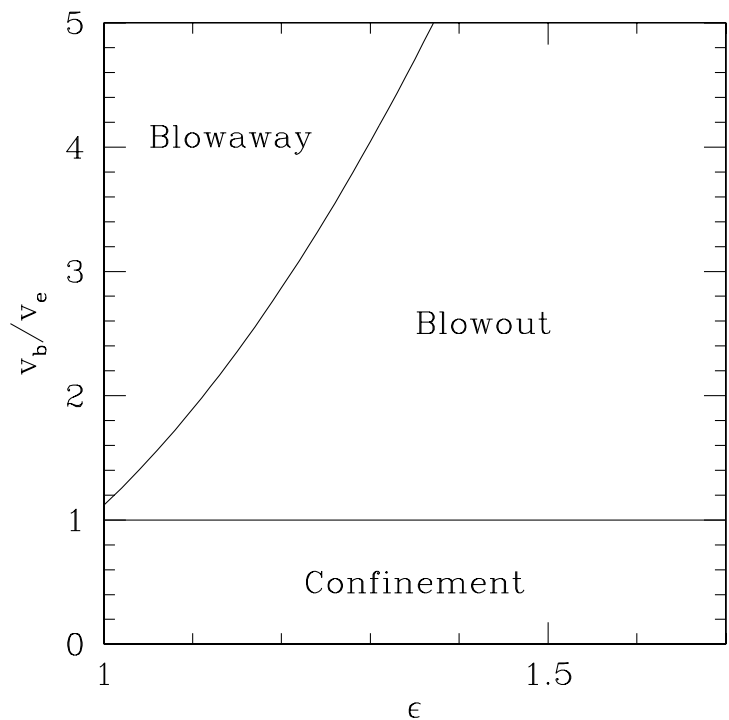}
\end{center}
\caption{Conditions for blowaway, 
blowout and confinement as a function of the major-to-minor axis ratio 
$\epsilon$ of dwarf galaxies; $\epsilon=1$ corresponds to spherical bodies.
This Figure is taken from Ferrara \& Tolstoy (2000).}
\label{fig:blowaway}
\end{figure}

\q
\frac{v_b}{v_e}\geq (\epsilon - a)^2 a^{-2} \mbox{e}^{3/2},
\nq

\noindent
where $a=2/3$.
The above equation is graphically displayed in Figure~\ref{fig:blowaway}. Flatter
galaxies (large $\epsilon$ values) preferentially undergo \index{blowout} blowout, whereas
rounder ones are more likely to be blown-away; as $v_b/v_e$ is increased the
critical value of $\epsilon$ increases accordingly. Unless the galaxy is 
perfectly spherical, \index{blowaway} blowaway is always preceded by blowout; between the
two events the aspect of the galaxy may look extremely perturbed, with one
or more huge cavities left after blowout.

\subsection{Further Model Improvements}

All models discussed so far make the so-called {\it SEX BOMB} assumption, i.e.
a Spherically EXpanding blastwave. This is a good assumption as long as
a single burst regions exists whose size is small compared to the size of the
system and it is located at the center of the galaxy. This however is a rather 
idealized situation.   
 
A more detailed study about the possibility of mass loss due to distributed {\index{supernova!explosion} SN explosions
at high redshift has been carried on by Mori, Ferrara \& Madau (2002). These authors
presented results from three--dimensional numerical simulations of the 
dynamics of SN--driven bubbles as they propagate through and escape the grasp 
of subgalactic halos with masses $M=10^8\,h^{-1}\,\Msun$ at redshift $z=9$. 
Halos in this mass
range are characterized by very short dynamical timescales (and even shorter
gas cooling times) and may therefore form stars in a rapid but
intense burst before SN `feedback' quenches further star formation.
This hydrodynamic simulations use a nested grid method to follow the evolution
of explosive multi--SN events operating on the characteristic timescale of a
few $\times 10^7\,$yr, the lifetime of massive stars. The results confirm
that, if the star formation efficiency of subgalactic halos is $\approx 10\%$,
a significant fraction of the halo gas will be lifted out of the potential
well \index{blowaway} (`blow--away'), shock the intergalactic medium, and pollute it with
metal--enriched material, a scenario recently advocated
by Madau, Ferrara, \& Rees (2001). 
Depending on the stellar distribution,
Mori et al. (2002) found that less than 30\% of the available SN energy gets 
converted into kinetic energy of the blown away material, the remainder
being radiated away. 

However, it appears that realistic models lead to the conclusion that 
mechanical feedback is less efficient than expected from SEX BOMB simple schemes. The 
reason is that off--nuclear \index{supernova!explosion} SN explosions drive inward--propagating \index{shock} shocks that tend to
collect and pile up cold gas in the central regions of the host halo.
Low--mass galaxies at early epochs then may survive multiple SN events and
continue forming stars.

Figure~\ref{fig:mori} and \ref{fig:12} show the fraction of the initial
halo baryonic mass contained inside (1, 0.5, 0.1) of the virial radius 
$r_{vir}$ as a function of time for two different runs: an extended stellar 
distribution (case~1) and a more concentrated one (case~2).
The differences are striking: in case~1, the amount of gas at the center is
constantly increasing, finally collecting inside 0.1$r_{vir}$ about 30\% of
the total initial mass. On the contrary, in case~2, the central regions 
remain practically devoided of gas until 60 Myr, when the accretion process
starts. The final result is a small core containing a fraction of only
5\% of the initial mass. In the former case, 50\% of the halo mass is ejected together with the
shell, whereas in case 2 this fraction is $\sim 85$\%, i.e., the \index{blowaway} blow-away is
nearly complete.

\begin{figure}
\begin{center}
\includegraphics[width=11cm]{sigrav_fig08.ps}
\end{center}
\caption{Snapshots of the logarithmic 
number density of the gas at five different elapsed times. The three panels 
in each row show the spatial density distribution in the $x-y$ plane on the 
nested grids. The density range is $-5\leq\log(n/\mbox{cm}^{-3})\leq 1$.
This Figure is taken from Mori et al. (2002)}
\label{fig:mori}
\end{figure}

\begin{figure}
\begin{center}
\includegraphics[width=8cm,angle=-90]{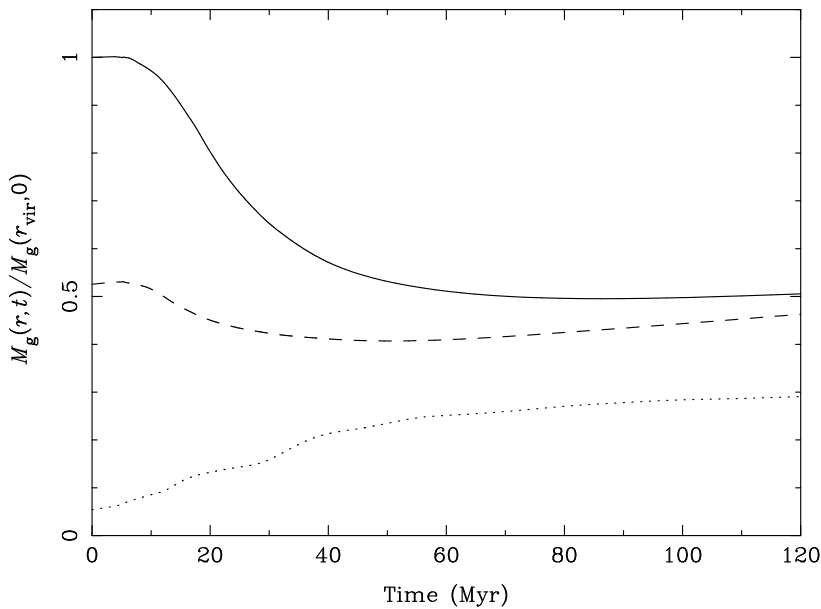}
\end{center}
\caption{The evolution of the gas mass inside the gravitational potential 
well of the CDM halo for an extended stellar distribution (case 1) 
as a function of time.
Curves correspond to the gas mass inside the virial radius 
$r_{\rm vir}$ ({\it solid line}), $0.5 r_{\rm vir}$ ({\it dashed line}), 
and $0.1 r_{\rm vir}$ ({\it dotted line}). This Figure is taken from Mori et
al. (2002).}
\label{fig:12}
\end{figure}

\begin{figure}
\begin{center}
\includegraphics[width=8cm,angle=-90]{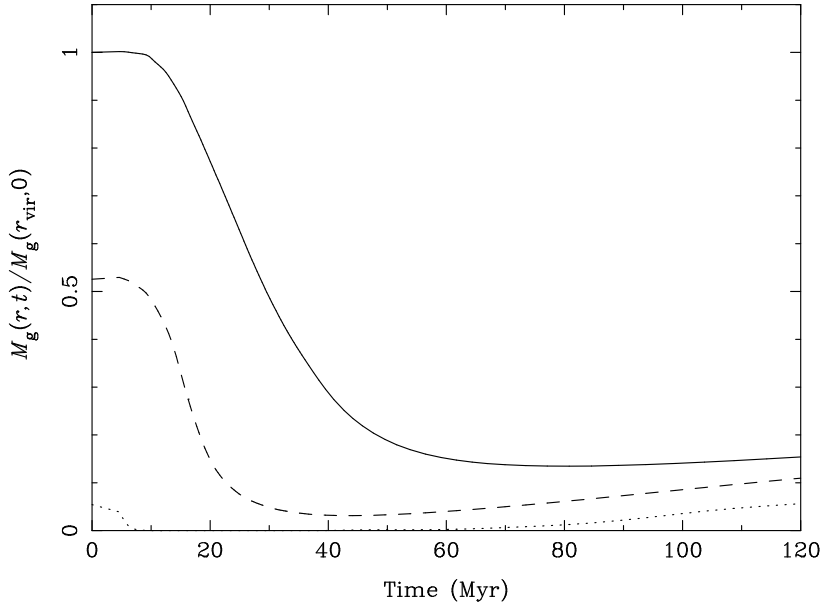}
\end{center}
\caption{The same of Figure~\ref{fig:12} but for a more concentrated stellar
distribution (case 2). This Figure is taken from Mori et al. (2002).}
\label{fig:14}
\end{figure}

As a first conclusion, it seems that quenching star formation in galaxies by
ejecting large fraction of their gas seems very difficult and hence unviable as
a feedback \index{feedback!scheme} scheme (although this might be possible in very small galaxies). 
Heavy elements can instead escape much more easily as they are carried away
by the mass-unloaded, hot, SN-shocked gas; energy is carried away efficiently 
as well. This conclusion raises the issue if star formation must be rather 
governed by more gentle and self-regulating processes, i.e. a `true' feedback.

\section{The Gentle Feedback}

In numerical simulations an astonishingly large number of attempts to implement
feedback \index{feedback!scheme} scheme have been tried. In spite of this wealth, essentially all
of them can be classified in one of the two following categories: (i) 
\index{feedback!thermal} `Thermal' feedback schemes in which the energy released by a supernova is assumed to 
simply heat the \index{ISM} ISM. The dynamics of the gas is only slightly affected: this is 
because the dense gas in star-forming region is able to radiate away this
heat input very quickly. Moreover it would imply a huge amount of hot gas, that
is hardly seen in local universe. (ii) \index{feedback!kinetic} `Kinetic' feedback schemes, in which the explosion energy 
is assigned to fluid parcels as pure kinetic energy. This algorithm performs slightly 
better and it is probably more physical. It would be impossible to review in detail here
even only a few of these schemes. We have thus chosen to describe a minimum feedback 
model that tries to catch the most important feature that is missing in the previous ones,
that is the multiphase structure and the matter exchange among gas phases characterizing
real galaxies.

\subsection{Feedback as \index{ISM} ISM Sterilization}

Dynamical models of the interstellar medium (ISM) describe this complex system 
in terms of a dense, cold neutral component (CNM) co-spatial with a 
warm neutral intercloud medium (WNM) by which it is pressure confined. 
Such a multiphase medium offers an efficient self-regulatory mechanism for
star formation. Schematically, as more stars 
are formed, the extra heating caused by the enhancement of their UV field tends
to transfer mass from the CNM into the WNM, where the conditions are highly unfavorable to 
star formation, i.e. a {\it sterile} phase. This acts to decrease the 
star formation rate and slowly brings back gas from the WNM reservoir in the cold phase
where stars can start to form again. Under most conditions this cycle is stable and tends to
regulate the amount of gas turned in stars quite effectively. 
In this scenario starburst 
can be only produced either by external triggers (mergings and interactions with other galaxies)  
and/or during the very first phase of galaxy formation, when the gentle feedback has not yet 
had the time to control the system. 
The physical basis of the gentle feedback is the \index{thermal instability} thermal instability already
discussed early on.  
Field (1965) discussed the criteria for stability of gas in thermal 
equilibrium (i.e. $L(n,T)=n\Gamma(n,T)-n^2\Lambda(n,T)=0$,
see also sect. \ref{sec:theq}). In a plot of the thermal
pressure $P/k$ versus the hydrogen density $n$, the region of thermal stability
occur for $d(\log P)/d(\log n)>0$. Thus, Figure~\ref{fig:Pn} shows that a
stable two-phase medium, can be maintained in pressure equilibrium for 
$P_{min}<P<P_{max}$. For pressure less than $P_{min}$ only the warm phase is 
possible, while at pressures greater than $P_{max}$ only the cold phase is 
possible (Wolfire et al. 1995). Moreover, in Figure~\ref{fig:Pn} it is shown,
that when the metallicity of the gas $Z$, assumed to be equal to the 
dust-to-gas ratio $D/G$, is decreased, the pressure range in which a 
multiphase medium can exist becomes wider; in addition, the mean equilibrium
pressure and density of the cold gas increase (Ricotti et al. 1997).


The first advantage of such feedback scenario is that there is no
need for gas loss as required by canonical feedback \index{feedback!scheme} schemes: typically,
the loss rate necessary to regulate the star formation activity is difficult 
to sustain for galaxies with mass of
the order of the Milky Way for interestingly long periods. In addition,
the production of large amounts of hot, X-ray emitting gas is expected.
Such gas is rarely seen in/around observed galaxies.   

An additional advantage of these models is the self-regulatory behavior of the
process, that we have seen to be typical of feedback. In fact, there is no 
need of infall or outflow of gas from the galaxy (close box). As cold gas is 
available, the galaxy has a star formation burst, leading to the increase of
the metallicity and of the dust content. Moreover, extra heating is provided
by the \index{supernova!explosion} SN explosions and from UV background, so that some gas mass is transferred
from the CNM to the WIM. The star formation is quenched since the gas is in the warm, 
{\it sterile} phase, and the heating drops down. After a cooling time, the 
gas becomes again available for star formation.

The disadvantage of these models is that they are physically complex, since
the involve the interplay between dynamics and thermodynamics. Hence, 
a number of physical processes await inclusion in current cosmological
simulations.

\begin{figure}
\begin{center}
\includegraphics[width=12cm]{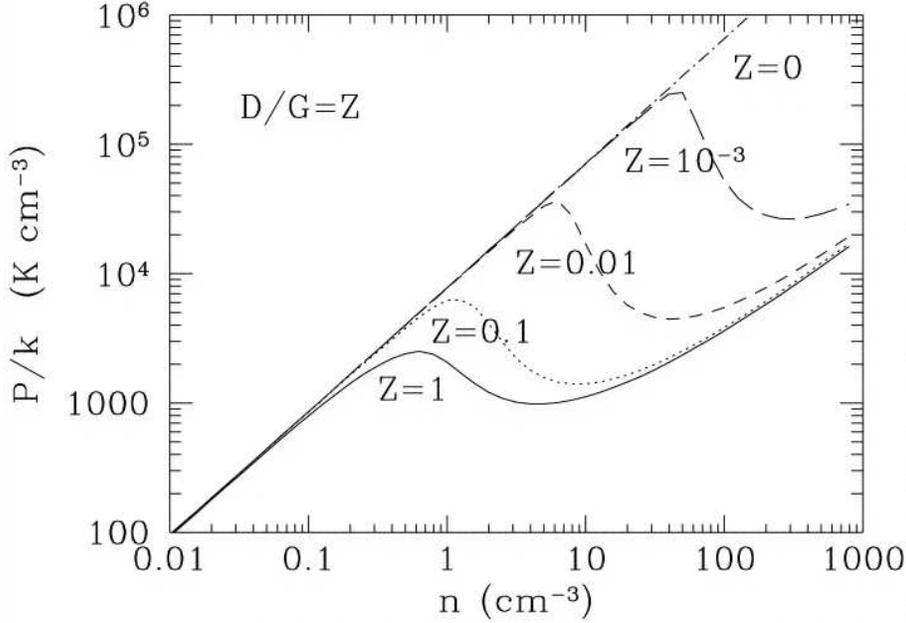}
\end{center}
\caption{Thermal pressure $P/k$ vs. hydrogen density
$n$; the curves refer to different values of the
dust-to-gas ratio, $D/G$, and metallicity $Z$, with $D/G
= Z$. The gas is thermally stable for $(d \log P/d \log n)>0$
(i.e. positive slope of the curves).
Unless $Z=0$, a stable two-phase medium is supported.
 This Figure is taken from Ricotti et al. (1997).}
\label{fig:Pn}
\end{figure}

\subsection{A \index{porosity} Porosity-regulated Feedback}

A similar feedback \index{feedback!scheme} scheme has been recently proposed by Silk (2003). In such
a feedback prescription, the filling factor of the hot, SN-shocked gas plays
a central role.
The filling factor $f_{hot}$ of hot gas can be expressed in terms of the
\index{porosity} porosity $Q$ as

\q
f_{hot}=1-\mbox{e}^{-Q},
\nq

\noindent
where $Q$ is regulated by the pressure confinement of the supernova bubbles.
As $f_{hot}\rightarrow 1$, the star formation rate reaches a saturation level
and for $f=1$ a `buoyant' wind is produced with outflow rate approximately
equal to the star formation rate. 

This feedback has the advantage to be based on the physical treatment of 
the \index{supernova!explosion} supernova explosion, that current numerical simulations fail to follow
in detail due to the lack of resolution. Moreover the expression of the
\index{porosity} porosity $Q$ is identical for all the masses: the outflow rate is of the order
of the star formation rate. Only the star formation efficiency depends on 
potential well depth ($\epsilon_{cr}\propto V_c^{2.7}$). The wind efficiency 
depends only weakly on $V_c$. In summary, outflows can occur even for massive
galaxies since star formation efficiency is larger   in these systems.

However, there are some problems. The \index{porosity} porosity-driven feedback is not a true multiphase
model and has to be generalized to include the time evolution of $Q$, 
metallicity, \index{UV background} UV background, etc. Moreover, it again implies a large amount of hot
gas that is  not detected in galaxies.

\subsection{Advanced Multiphase/Feedback \index{feedback!multiphase scheme} Schemes}

The main problem affecting numerical simulations
of feedback processes is that, owing to the limited numerical resolution, 
there is  a large disparity between the 
minimum simulation timescale and the true physical timescale associated with
supernova feedback (Thacker \& Couchman 2001). 
Physically, following a supernova event, the gas cooling 
time very rapidly increases to $O(10^7)$ yr over the time it takes the \index{shock} shock 
front to propagate. Thus, it is the timescale - $t_{\dot{E}}$ - for the cooling
time to increase markedly that is of critical importance. 
In the SPH simulations, 
`stars' form and `supernova feedback' occurs in gas cores that are typically 
less than $h_{min}/5$ in diameter ($h_{min}$ being the effective spatial 
resolution). Any rearrangement of the particles in such a small region leads 
to a density estimate varying by at most $\sim 7$\%. Since the cooling time in 
the model is dependent on the local SPH density, any energy deposited into a 
feedback region only reduces the local cooling rate by increasing the 
temperature; the SPH density cannot respond on a timescale that in real 
supernova events would very rapidly increase the cooling time. A successful 
feedback algorithm in an SPH model must thus overcome the fact that 
$\rho_{SPH}$ does not change on the same timescale as $t_{\dot{E}}$ and
consequently that cooling times for dense gas cores remain short.

Standard implementations of SPH overestimate the density for particles
that fall near the boundary of a higher density phase
(Pearce et al. 1999). The usual assumption of small density gradients
across the smoothing kernel breaks down in this regime, and nearby
clusters of high density particles cause an upward bias in the standard SPH 
estimate:

\begin{equation}
<\rho_{i}>=\sum_{j}^{N} m_{j} W_{ij},
\label{std_dens_eq}
\end{equation}

\noindent
where $N$ is the number of neighbors $j$ of particle $i$  and $W_{ij}$
is a symmetric smoothing kernel. In order to avoid
this bias, which leads to artificially high cooling rates and
to spatial exclusion effects, the neighbor search and the
density evaluation of equation (\ref{std_dens_eq}) has to be modified in a 
way which leaves the numerical scheme as simple as possible.
It is important to recognize that the local density is the gas property
responsible for phase segregation (since it determines the local
cooling rate). An appropriate density estimator for a particle of any
given phase should use only local material which is also part of that
same phase. 

Marri \& White (2003) proposed a new \index{feedback!multiphase scheme} feedback scheme that takes into account 
these considerations. This works as follows. As a supernova explodes, 
the energy is distributed in a fraction $\epsilon_c$ ($\epsilon_w$) to the 
cold (warm) gas. The fundamental
hypothesis is that at large scale (of kpc order, say) the net effect of all
the complex {\it microscopic} processes is well described by an energy input
shared by the {\it macroscopic} phases in given proportions. Values for 
$\epsilon_c$ and $\epsilon_w$ could be fixed from a complete theory of the
\index{ISM}
ISM, describing all the relevant processes, or through numerical simulations.
Feedback to the hot phase is implemented by adding thermal energy to the ten
nearest hot neighbors. Feedback to the cold phase is instead accumulated in
a reservoir within the star-forming particle itself. This continues until the
accumulated energy is sufficient to heat the gas component of the particle 
above $T\sim 50 T_\star\sim 10^6$ K. This is far enough above $T_\star$ for
the promoted particle to be considered `hot' in its subsequent evolution. 
At the same time any `hidden' stellar content is dumped to the cold neighbors.

A simple test of these new \index{feedback!multiphase scheme} scheme (the initial condition are the rotating, 
centrally concentrated sphere described in Navarro \& White (1993) which
consists of 90\% by mass in dark matter) shows that hot particle do survive 
near the central disk in the multiphase SPH case. They have an almost 
spherically symmetric distribution with density peaked at the center (Figure 
\ref{fig:position}). The 
cold disk rotates within this ambient hot medium. In contrast, in the standard
SPH model hot particle are excluded from the vicinity of the disk so that the
hot phase actually has a density minimum at the center of the galaxy. This
is seen most clearly in the gas density profile of Figure \ref{fig:prof}. Such
behavior is clearly unphysical.

\begin{figure}
\begin{center}
\includegraphics[width=12cm]{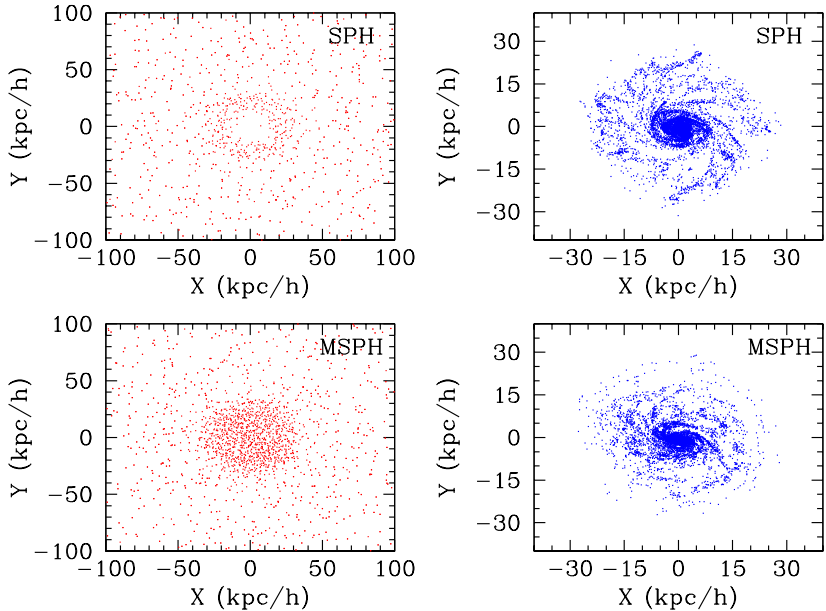}
\end{center}
\caption{Face-on projections at
$t\sim 1.2$ Gyr of 8000 gas particle, cooling only simulations
of a collapsing, rotating sphere. The standard SPH
simulation is in the upper row with the MSPH simulation below it.
The left-hand plots show ``hot'' particles ($T> 10^5$K) while the
right hand ones show ``cold'' particles ($T< 10^5$K and $n_H>0.1$
cm$^{-3}$). This Figure is taken from Marri \& White (2003).}
\label{fig:position}
\end{figure}

\begin{figure}
\begin{center}
\includegraphics[width=12cm]{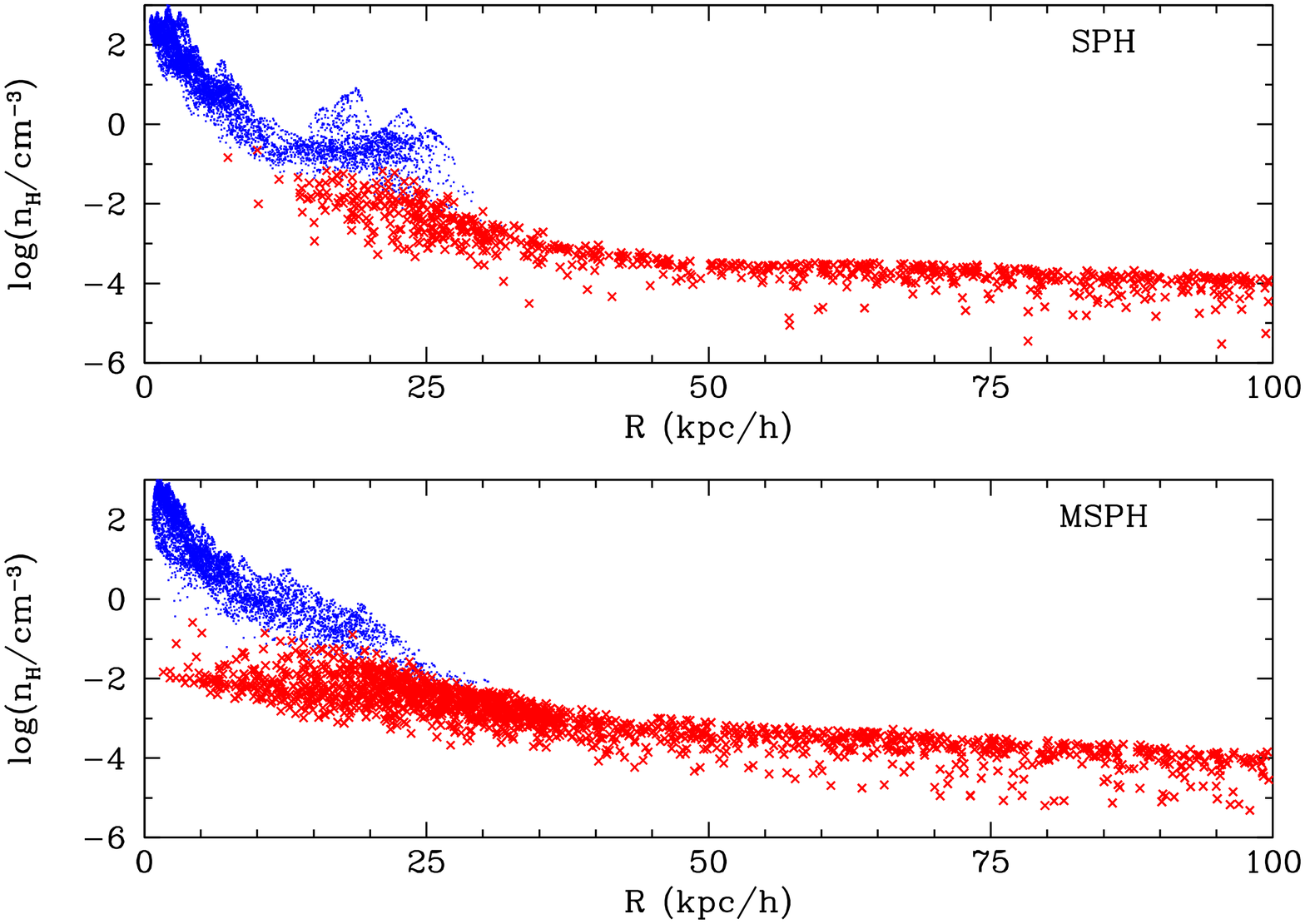}
\end{center}
\caption{Particle densities as a function of
galactocentric distance after $\sim 1.2$ Gyr of evolution in two 8000
gas particle, cooling only simulations of the collapse of a rotating
sphere. The upper plot is for an SPH model and the lower for an MSPH
model. Gas particles with $T>10^5$K are plotted with crosses. This Figure
is taken from Marri \& White (2003).}
\label{fig:prof}
\end{figure}

\section{Feedbacks in the Early Universe}

At $z=1000$ the Universe has cooled down to 3000 K and the hydrogen becomes
neutral (recombination). Then, at $z\simlt 20$ the \index{Pop III} first stars (Population 
or Pop III stars) form and these gradually photo-ionize the hydrogen in the
\index{IGM} IGM (reionization). These epoch, witnessing the return of light in the universe
after the Big Bang, is usually dubbed as the {\it Cosmic Dawn}.  
At $z<6$, galaxies form most of their stars and grow by
merging. Finally at $z<1$, the massive clusters are assembled.

We can distinguish three main feedback types that are fundamentally
shaping the universe at cosmic dawn: 

\begin{enumerate}
\item {\bf Stellar Feedback}: \index{feedback!stellar} the process of star formation produces the
conditions (destruction of cold gas) such that astration (or even galaxy
formation) comes to a halt or is temporarily blocked;
\item{\bf Chemical Feedback}: \index{feedback!chemical} massive stars by polluting the ambient gas with
heavies, shift the prevailing formation mode toward low mass stars 
(self-killing population)
\item{\bf Radiative Feedback}: \index{feedback!radiative} the process of star formation erases the 
conditions (molecular formation) necessary to form stars via UV radiation and
ionizing photons. Such radiative input heats and ionizes the gas decreasing its 
ability to produce sources (stars) to further sustain reionization.
\end{enumerate}
In the following we will describe in detail the three feedback types.
\subsection{Stellar Feedback}

\subsubsection{Oxygen in the Low Density \index{IGM} IGM}

One of the most obvious signatures of \index{feedback!stellar} stellar feedback is the metal enrichment
of the intergalactic medium. If powerful winds are driven by \index{supernova!explosion} supernova explosions,
one would expect to see widespread traces of heavy elements away from their
production sites, i.e. galaxies. 
Schaye et al. (2000) have reported the detection of OVI in the low-density
\index{IGM} IGM at high redshift. They perform a pixel-by-pixel search for OVI 
\index{absorption} absorption
in eight high quality quasar spectra spanning the redshift range $z=2.0-4.5$.
At $2\simlt z\simlt 3$ they detect OVI in the form of a positive correlation
between the HI Ly$\alpha$ \index{optical depth} optical depth and the optical depth in the
corresponding OVI pixel, down to $\tau_{\rm HI}\sim 0.1$ (underdense regions),
that means that metals are far from galaxies (Figure~\ref{fig:ovi}). 
Moreover, the observed narrow widths of metal \index{absorption} absorption lines (CIV, SiIV)
lines lines imply low temperatures $T_{\rm \index{IGM} IGM}\sim {\rm few}\times 10^4$ K.

A natural hypothesis would be that the \index{Ly$\alpha $!forest} Ly$\alpha$ forest has been enriched by metals ejected
by Lyman Break Galaxies at moderate redshift. The density of these objects is
$n_{\rm LBG}=0.013$ $h^3$ Mpc$^3$. A filling factor of $\sim 1$\% is obtained
for a \index{shock} shock radius $R_s=140$ $h^{-1}$ kpc, that corresponds at $z=3$ ($h=0.5$) 
to a \index{shock} shock velocity $v_s=600$ km s$^{-1}$. In this case, we expect a 
post\index{shock} shock gas temperature larger than $2\times 10^6$ K, that is around hundred times what
we observed. So the metal pollution must have occurred earlier than redshift 3,
resulting in a more uniform distribution and thus enriching vast regions of 
the intergalactic space. This allows the \index{Ly$\alpha $!forest} Ly$\alpha$ forest to be hydrodynamically
`cold' at low redshift, as intergalactic baryons have enough time to relax again
under the influence of dark matter gravity only (Scannapieco, Ferrara \& Madau
2002).

\bigskip
In Figure~\ref{fig:tigm} is shown the \index{thermal history} thermal history of the \index{IGM} IGM                    
as a function of redshift as computed by Madau, Ferrara \& Rees (2001). 
The gas is allowed to interact with the CMB 
through Compton cooling and either with a time-dependent quasar-ionizing 
background as computed by Haardt \& Madau (1996) or with a time-dependent
metagalactic flux of intensity 
$10^{-22}\,{\rm\,erg\,cm^{-2}\,s^{-1}\,Hz^{-1}\,sr^{-1}}$ at 1 Ryd and 
power--law spectrum with energy slope $\alpha=1$. The temperature of the 
medium at $z=9$ has been either computed self-consistently from 
photoheating or fixed to be in the range $10^{4.6}-10^5$ K, as expected in   
SN-driven bubbles with significant filling factors. The various curves show
that the temperature of the \index{IGM} IGM at $z=3-4$ will retain little memory of an
early era of pregalactic outflows.

\bigskip
The large increase of the \index{IGM} IGM temperature at high redshift 
connected with such an era of pregalactic outflows, causes a larger 
Jeans mass, thereby preventing gas from accreting efficiently
into small dark matter halos (Benson \& Madau 2003). 
For typical preheating energies,
the \index{IGM} IGM is driven to temperatures just below the virial 
temperature of halos hosting $L_\star$ galaxies. Thus we may expect preheating
to have a strong effect on the galaxy luminosity function at $z=0$.
Moderate preheating scenarios, with $T_{\rm \index{IGM} IGM}\simgt 10^5$ K at $z\sim 10$,
are able to flatten the faint-end slope of the luminosity function, producing
excellent agreement with observations, without the need for any local 
strong feedback within galaxies (Figure \ref{fig:LF}).  

\begin{figure}
\begin{center}
\includegraphics[width=11cm]{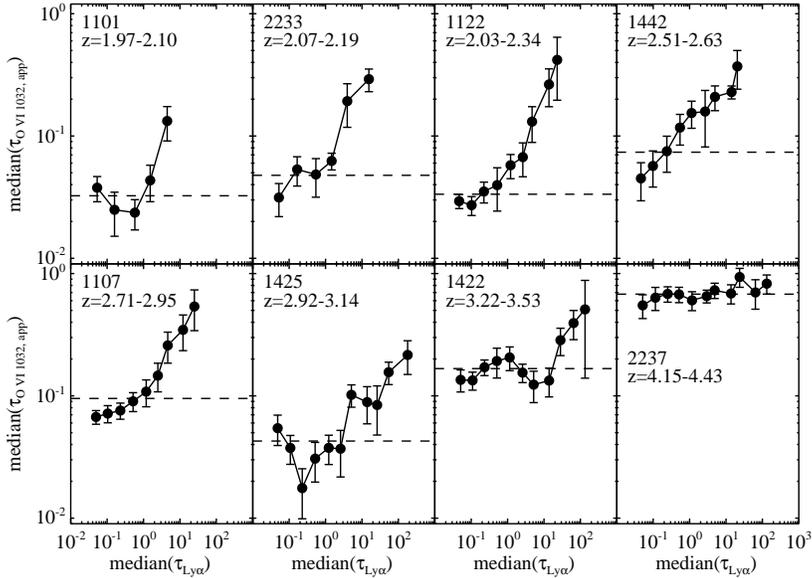}
\end{center}
\caption{IGM Ly$\alpha$-OVI optical depth correlation from the pixel analysis
of the spectra of 8 QSOs. Vertical error bars are $1\sigma$ errors,
horizontal error bars are smaller than the symbols and are not
shown. For $z\simlt 3$ $\tau_{\rm OVI, app}$ and $\tau_{\rm HI}$ are
clearly correlated, down to optical depths as low as $\tau_{\rm HI}
\sim 10^{-1}$. A correlation between $\tau_{\rm OVI, app}$ and
$\tau_{\rm HI}$ implies that OVI absorption has been detected in the
\index{Ly$\alpha$!forest} Ly$\alpha$ forest (Schaye et al. 2000).}
\label{fig:ovi}
\end{figure}

\begin{figure}
\begin{center}
\includegraphics[width=12cm]{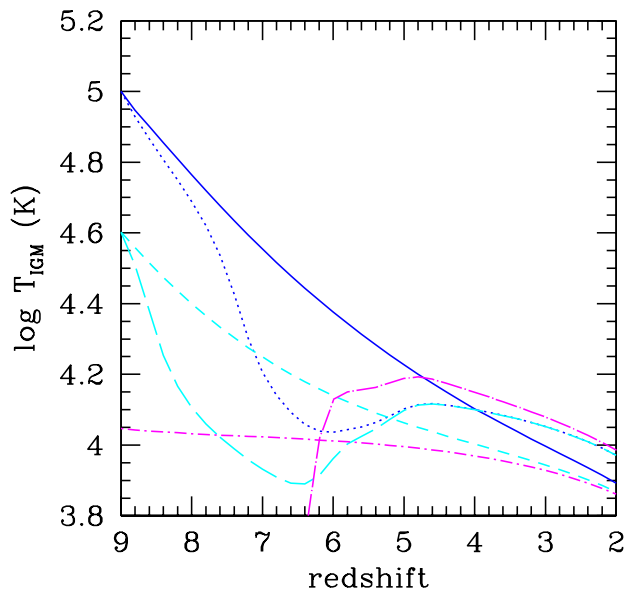}
\end{center}
\caption{\index{thermal history} Thermal history of intergalactic gas at the mean density in an
Einstein--de Sitter universe with $\Omega_bh^2=0.019$ and $h=0.5$.
{\it Short dash--dotted line:} temperature evolution when the
only heating source is a constant \index{UV background} ultraviolet (CUV) background of intensity
$10^{-22}\,{\rm\,erg\,cm^{-2}\,s^{-1}\,Hz^{-1}\,sr^{-1}}$
at 1 Ryd and power--law spectrum with energy slope $\alpha=1$.
{\it Long dash--dotted line:} same for the time--dependent quasar ionizing
background as computed by Haardt \& Madau (1996; HM).
{\it Short dashed line:} heating due to a CUV background  but with an initial
temperature of $4\times 10^4\,$K at $z=9$ as expected from an early era
of pregalactic outflows.
{\it Long dashed line:} same but for a HM background. {\it Solid line:} heating
due to a CUV background  but with an initial temperature of $10^5\,$K
at $z=9$.
{\it Dotted line:} same but for a HM background. This Figure is taken from
Madau, Ferrara \& Rees (2001)}
\label{fig:tigm}
\end{figure}

\begin{figure}
\begin{center}
\includegraphics[width=12cm]{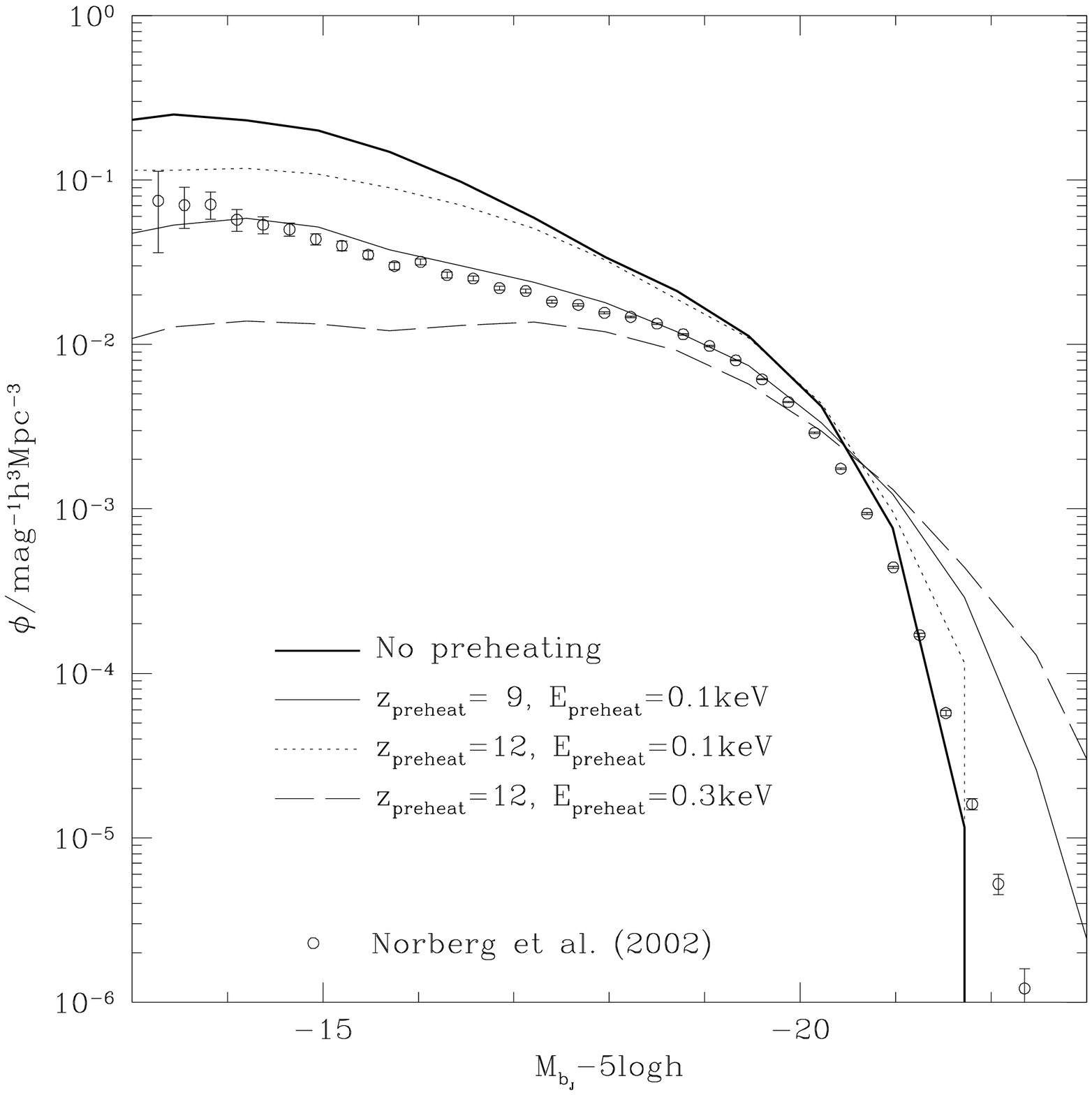}
\end{center}
\caption{B-band luminosity functions of galaxies at $z=0$, as
predicted by the semi-analytic model of Benson et al. (2002). 
The observational determination of Norberg et al. (2002) is shown as circles. 
This Figure is taken from Benson \& Madau (2003).}
\label{fig:LF}
\end{figure}

\subsubsection{\index{gas stripping} Gas Stripping from Neighbor Galaxy \index{shock} Shocks}

The formation of a galaxy can be inhibited also by the outflows from neighboring
\index{dwarf galaxy} dwarfs as the result of two different mechanisms (Scannapieco et al. 2000).
In the \index{mechanical evaporation} ``mechanical evaporation'' scenario, the gas associated
with an overdense region is heated by a \index{shock} shock above its 
virial temperature. The thermal pressure of the gas then
overcomes the dark matter potential and the gas expands out of the
halo, preventing galaxy formation.  In this case, the cooling
time of the collapsing cloud must be shorter than its sound crossing
time, otherwise the gas will cool before it expands out of the gravitational 
well and will continue to collapse.

Alternatively, the gas may be stripped from a collapsing perturbation by a \index{shock} shock
from a nearby source.  In this case, the momentum of the 
\index{shock} shock is sufficient to carry with it the gas associated with the
neighbor, thus emptying the halo of its baryons and preventing a galaxy from 
forming.

In principle, outflows can suppress the formation of nearby galaxies both by
\index{shock} shock heating/evaporation and by \index{gas stripping} stripping of the baryonic matter
from collapsing dark matter halos; in practice, the short cooling times for 
most dwarf-scale collapsing objects suggest that the baryonic stripping 
scenario is almost always dominant (Figure~\ref{fig:scanna}).  
This mechanism has the largest impact in
forming \index{dwarf galaxy} dwarfs in the $\simlt 10^9 M_\odot$ range which is sufficiently
large to resist photoevaporation by UV radiation, but too small to
avoid being swept up by nearby dwarf outflows. 

\begin{figure}
\begin{center}
\includegraphics[width=12cm]{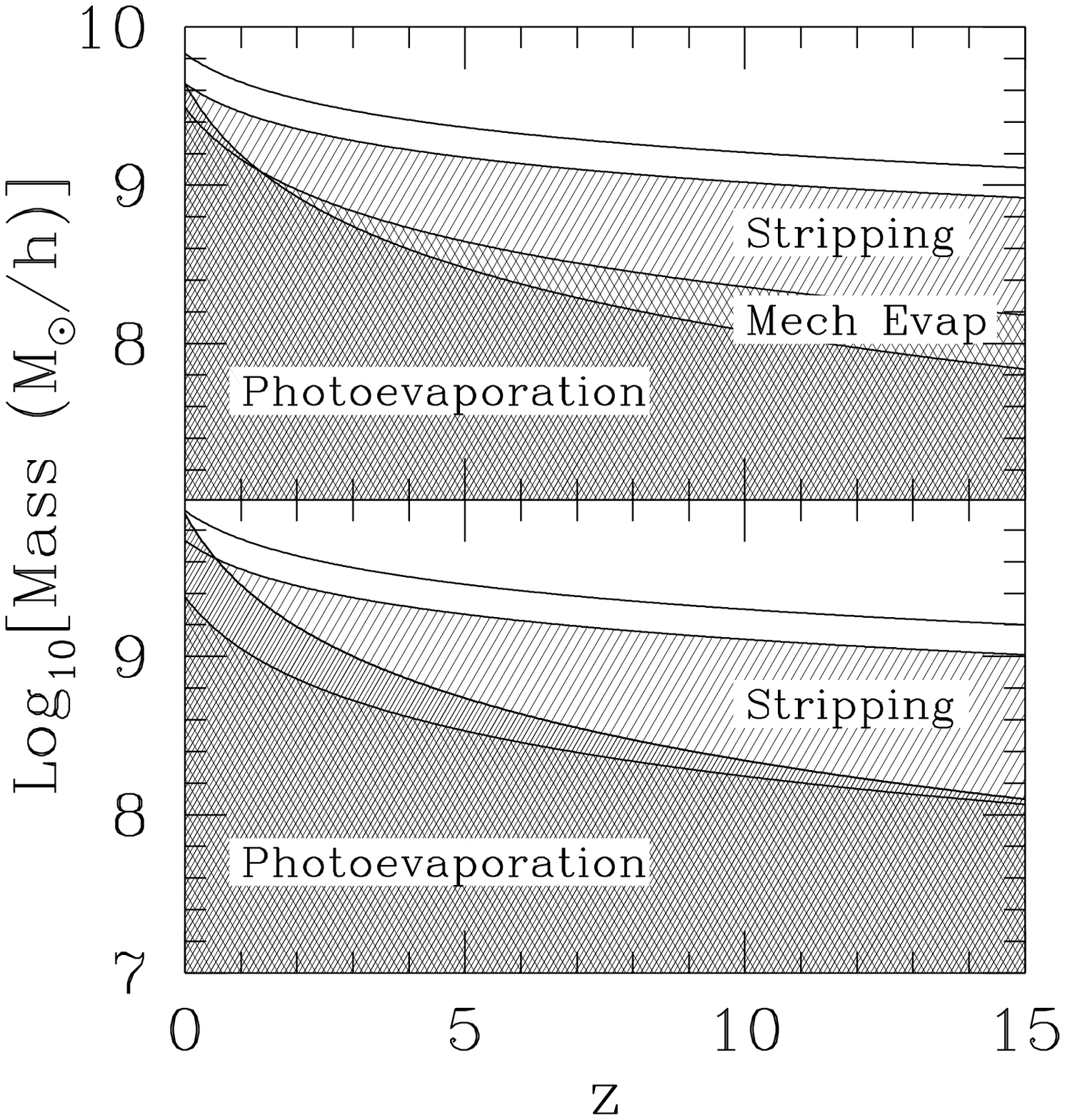}
\end{center}
\caption{Relevant mass scales for suppression of dwarf galaxy formation.
The upper lines are the masses
below which halos will be heated beyond their virial temperatures,
although cooling prevents \index{mechanical evaporation} mechanical evaporation from occurring for halos
with masses above the cross hatched regions.  The second highest set of
lines, bounding the lightly shaded regions, show the masses below which
baryonic \index{gas stripping} stripping is effective.   Finally the 
heavily shaded regions show objects that are susceptible to photoevaporation.
The upper panel 
is a flat CDM model and the lower panel is a flat $\Lambda$CDM model
with $\Omega_0 = 0.3$.  In all cases $(\epsilon N h) = 5000 \Omega_0^{-1}$,
the overdensity $\delta = \rho/\rho_0 = 2.0$, $\Omega_b = 0.05$, $h = .65$.
Note that photoevaporation affects a larger mass range than mechanical
evaporation in the $\Lambda$CDM cosmology. This Figure is taken from
Scannapieco et al. (2000)}
\label{fig:scanna}
\end{figure}

\subsection{\index{feedback!chemical} Chemical Feedback}

Recent studies (Bromm, Coppi \& Larson 1999, 2002; Abel, Bryan 
\& Norman 2000; Schneder et al. 2002; Omukai \& Inutsuka 2002; Omukai \& Palla
2003) have shown that in absence of heavy elements the formation 
of  stars with masses 100 times that of the Sun would have been strongly 
favored, and that low-mass stars could not have formed before a
minimum level of metal enrichment had been reached.

Bromm et al. (2001) have studied the effect of the \index{metallicity} metallicity on the evolution
of the gas in a collapsing dark matter mini-halo, simulating an isolated 
3$\sigma$ peak of mass $2\times10^6\;\Msun$ that collapse at $z\sim 30$,
using smoothed particle hydrodynamics. The gas has a supposed level of 
\index{pre-enrichment} pre-enrichment of either $Z=10^{-4}\;\Zsun$ of $10^{-3}\;\Zsun$. The H$_2$ is
assumed to be radiatively destroyed by the presence of a \index{UV background!soft} soft UV background.
Moreover Bromm et al. (2001) do not consider the presence of molecules or
dust.

The evolution proceeds very differently for the two cases. The gas in the
lower \index{metallicity} metallicity case fails to undergo continued collapse and fragmentation,
remaining in the post-virialization, pressure-supported state in a roughly
spherical configuration. The final result is likely to be the formation of
a single very massive ($\sim 10^3-10^4\;\Msun$) object (Figure \ref{fig:runA}).
On the other hand (Figure \ref{fig:runB}),
the gas in the higher \index{metallicity} metallicity case can cool efficiently and collapse into
a disk-like configuration. The disk material is gravitational unstable and 
fragments into a large number of high density clumps, that are the seeds of
low mass stars. So, they conclude that there exists a \index{metallicity!critical} critical metallicity
$Z_{crit}$ at which the transition from the formation of high mass 
objects to a low-mass star formation mode, as we observe at present, occurs.
In the following we refer to \index{Pop III} Pop III stars in the mass range 100-600 $\Msun$
forming out of the collapse of $Z<Z_{crit}$ gas clouds.

\begin{figure}
\begin{center}
\includegraphics[width=12cm]{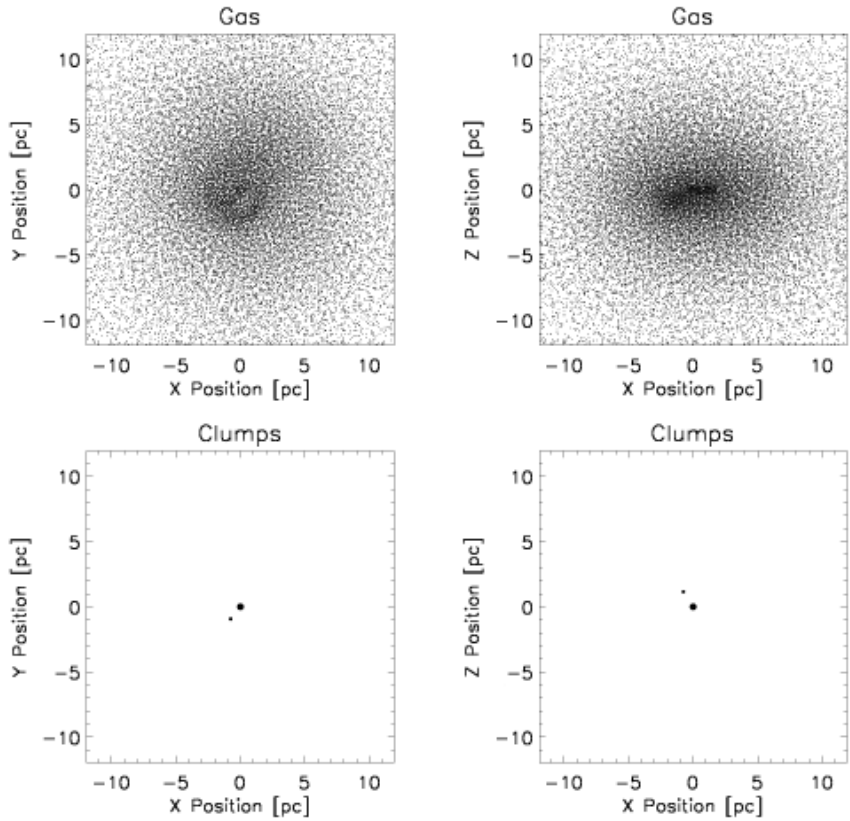}
\end{center}
\caption{Simulation of a collapsing dark 
matter halo with mean metallicity of $Z=10^{-4}\;\Zsun$. Morphology at 
$z=22.7$. Top row: the remaining gas in the diffuse phase. Bottom row: 
distribution of clumps. Dot sizes are proportional to the mass of the clumps.
Left panels: face-on view. Right panels: edge-on view. The box size is 30 pc.
It can be seen that no fragmentation occurs. This Figure is taken from
Bromm et al. (2001).}
\label{fig:runA}
\end{figure}

\begin{figure}
\begin{center}
\includegraphics[width=12cm]{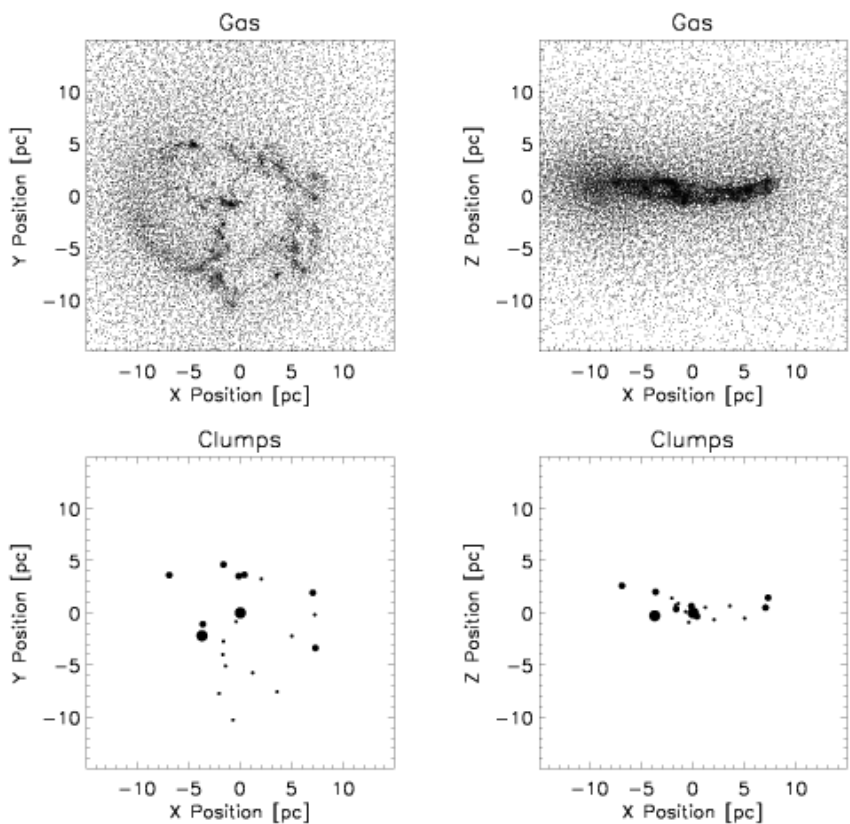}
\end{center}
\caption{Simulation of a collapsing dark 
matter halo with mean metallicity of $Z=10^{-3}\;\Zsun$. Morphology at
$z=28$. The convention of Figure \ref{fig:runA} is adopted for the row and
columns. It can be seen that the gas has undergone vigorous fragmentation.
This Figure is taken from Bromm et al. (2001)}
\label{fig:runB}
\end{figure}

More recently, Schneider et al. (2002) have investigated the problem more in
detail including the cooling due to H$_2$, molecular, and dust, using the
model of Omukai (2001). The gas within the dark matter halo is given an initial
temperature of 100 K, and the subsequent thermal and chemical evolution of the
gravitationally collapsing cloud is followed numerically until a central 
protostellar core forms. The results for different \index{metallicity} metallicities are summarized
in Figure \ref{fig:schneider1}, showing the temperature and adiabatic index 
evolution as a function of the hydrogen number density of protostellar clouds
for different metallicity ($Z=0,10^{-6},10^{-4},10^{-2},1\;\Zsun$).
The necessary conditions to stop fragmentation and start gravitational 
contraction within each clump are that cooling becomes inefficient, and the 
Jeans mass of the fragments does not decrease any further, thus favoring 
fragmentation into sub-clumps. This  condition depends somewhat on the 
geometry of the fragments, and translates into $\gamma >4/3$ ($\gamma > 1$) 
for spherical (filamentary) clumps, where $\gamma$ is the adiabatic index 
defined as $T \propto n^{\gamma-1}$. 

For a metal-free gas, the only efficient coolant is \index{molecular hydrogen} 
 molecular hydrogen. Cooling
due to molecular line emission becomes inefficient at densities above 
$n > 10^3 \,\mbox{cm}^{-3}$, and fragmentation stops when the minimum
fragment mass is of order $10^3 -10^{4}\;\Msun$. Clouds with mean \index{metallicity} metallicity 
$Z=10^{-6}\;\Zsun$ follow the same evolution as that of the gas with 
primordial composition in the $(n,T)$ plane. For metallicities larger than
$10^{-4}\;\Zsun$ the fragmentation proceeds further until the density is
$\sim 10^{13}$ cm$^{-3}$ and the corresponding Jeans mass is of the order
$10^{-2}\;\Msun$. Schneider et al. (2002) concluded that the 
\index{metallicity!critical} critical 
metallicity locates in the range $10^{-6}-10^{-4}\;\Zsun$.

According to the scenario proposed above, the first stars that form out of 
gas of primordial composition tend to be very massive, with masses 
$\sim 10^2-10^3\;\Msun$. It is only when metals change the composition of
the gas that further fragmentation occurs, producing stars with significantly
lower masses. A series of numerical studies (Heger \& Woosley 2001; Fryer et
al. 2001; Umeda \& Nomoto 2002) have investigated the nucleosynthesis and
final state of metal-free massive stars. Heger \& Woosley (2001) delineate
three mass ranges characterized by distinct evolutionary paths:

\begin{enumerate}
\item $M_\star\simgt 260\;\Msun$. -- The nuclear energy released from the 
collapse of stars in this mass range is insufficient to reverse the implosion.
The final result is a very massive black hole (VMBH) locking up all heavy 
elements produced.
\item $140\;\Msun\simlt M_\star\simlt 260\;\Msun$. -- The mass regime of the
\index{supernova!pair-unstable} pair-unstable supernovae (SN$_{\gamma\gamma}$). Precollapse winds and
pulsations result in little mass loss; the star implodes to a maximum 
temperature that depends on its mass and then explodes, leaving no remnant.
The explosion expels metals into the surrounding ambient \index{ISM} ISM.
\item $30\;\Msun\simlt M_\star\simlt 140\;\Msun$. -- Black hole formation
is the most likely outcome, because either a successful outgoing shock fails
to occur or the shock is so weak that the fallback converts the neutron star
remnant into a black hole (Fryer 1999)
\end{enumerate}

Stars that form in the mass ranges 1 and 3 above fail to eject most of their
heavy elements. If the first stars have masses in excess of 260 $\Msun$, they
invariably end their lives as VMBHs and do not release any of their 
synthesized heavy elements. However, as long as the gas remains metal
free, the subsequent generations of stars will continue to be top-heavy.
This `star formation conundrum' can be solved only if a fraction 
$f_{\gamma\gamma}$ of the first generation of massive stars are in the 
\index{supernova!pair-unstable}
SN$_{\gamma\gamma}$ range and enrich the gas with heavy elements up to a 
mean metallicity of $Z\geq 10^{-5\pm1}\;\Zsun$.

Schneider et al. (2002) computed the transition redshift $z_f$ at
which the mean \index{metallicity} metallicity is $10^{-4}\;\Zsun$ for various values of 
$f_{\gamma\gamma}$. The results are plotted in Figure \ref{fig:schneider2} along
with the corresponding critical density $\Omega_{VMBH}$ contributed by the
VMBHs formed. In order to set some limits on $f_{\gamma\gamma}$, we have to 
compare the predicted critical density of VMBH remnants to present 
observational data. This depends on the assumptions about the fate of these
VMBHs at late times.

Under the hypothesis that all VMBHs have, during the course of galaxy mergers,
been used to build up the supermassive black holes (SMBHs) detected today 
so that $\Omega_{VMBH}=\Omega_{SMBH}$, from the top panel of Figure
\ref{fig:schneider2} we can derive $f_{\gamma\gamma}\simeq 0.06$ and
$z_f\simeq 18.5$. 

\begin{figure}\begin{center}
\includegraphics[width=10cm]{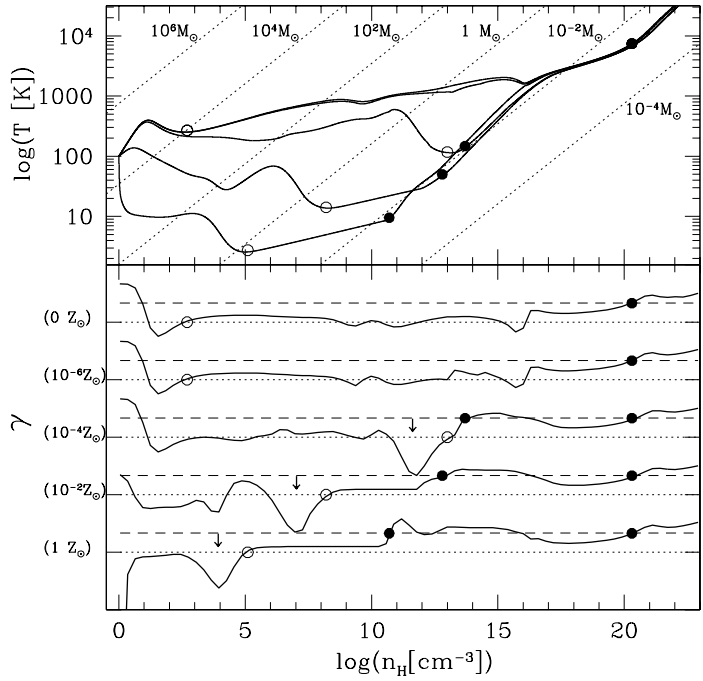}
\end{center}
\caption{{\it Top}: Evolution of the 
temperature as a function of the hydrogen number density of protostellar 
clouds with the same initial gas temperature but varying the metallicities
$Z=(0,10^{-6},10^{-4},10^{-2},1)\;\Zsun$ ($Z$ increasing from top to bottom
curves). The dashed lines correspond to the constant Jeans mass for spherical
clumps; open circles indicates the points where fragmentation stops; filled
circles mark the formation of hydrostatic cores. {\it Bottom}: The adiabatic
index $\gamma$ as a function of the hydrogen number density for the curves
shown in the top panel. Dotted (dashed) lines correspond to $\gamma=1$ 
($\gamma=4/3$); open and filled circles as above. This Figure is taken from
Schneider et al. (2002).}
\label{fig:schneider1}
\end{figure}

At the other extreme, wherein the assembled SMBHs in galactic centers have 
formed primarily via accretion and are unrelated to VMBHs, two possibilities
can be distinguish: (1) VMBHs would still be in the process of spiraling into 
the center of galaxies because of dynamical friction but are unlikely to have
reached the center within a Hubble time because of the long dynamical friction 
timescale (Madau \& Rees 2001), or, (2) VMBHs contribute the entire
baryonic dark matter in galactic halos.
In case (1) some fraction of these VMBHs might appear as off-center accreting
sources that show up in hard X-ray wave band. In fact, {\it ROSAT} and 
{\it Chandra} have detected such objects. Using the contribution to $\Omega$
from \index{X-ray {\it  ROSAT} source} X-ray-bright, off-center {\it ROSAT} sources as a limit for the density of
VMBHs, from the lower dashed line in the bottom panel of 
Figure \ref{fig:schneider2} $f_{\gamma\gamma}$ has to be nearly unity and 
$z_f\simgt 22.1$.
In case (2), assuming that the baryonic dark matter in galaxy halos is entirely
contributed by VMBH (upper dashed line in the {\it bottom panel} of 
Figure \ref{fig:schneider2}), $f_{\gamma\gamma}\simeq 3.15\times10^{-5}$ and
$z_f\simgt 5.4$. Therefore the case in which SMBHs are unrelated to VMBHs, 
only weak limits for $f_{\gamma\gamma}$ and $z_f$ can be obtained.
Although the actual data do not allow us at present to strongly constrain 
these two quantities, they provide interesting bounds on the proposed scenario.

\bigskip
We have seen that there exists a \index{metallicity!critical} critical metallicity 
$Z_{cr}=10^{5\pm1}\;\Zsun$ (Schneider et al. 2002; 2003) at which we 
expect the transition between a \index{IMF!top-heavy} top-heavy and a normal \index{IMF!Salpeter}
Salpeter-like IMF.
One point that might have important observational consequences is the fact
that cosmic metal enrichment has proceeded very inhomogeneously (Scannapieco,
Ferrara \& Madau 2002; Furlanetto \& Loeb 2003; Marri et al. 2003), with 
regions close to star formation sites rapidly becoming metal-polluted and
overshooting $Z_{cr}$, and others remaining essentially metal-free. Thus,
the two modes of star formation, \index{Pop III} Pop III and normal, must have been active 
at the same time and possibly down to relatively low redshifts, opening up
the possibility of detecting \index{Pop III} Pop III stars. 

Scannapieco, Schneider \& Ferrara (2003) have studied using an analytic model
of inhomogeneous structure formation, the evolution of Pop III objects as a
function of the star formation efficiency, IMF, and efficiency of outflow
generation. They parametrized the \index{feedback!chemical} chemical feedback through a single quantity,
$E_g$, which represents the kinetic energy input per unit gas mass of
outflows from \index{Pop III} Pop III galaxies. This quantity is related to the number
of exploding \index{Pop III} Pop III stars and therefore encodes the dependence on the
assumed IMF. For all values of the feedback parameter, $E_g$, Scannapieco
et al. (2003) found that, while the peak of \index{Pop III} Pop III star formation occurs at
$z\sim 10$, such stars continue to contribute appreciably to the star formation
rate density at much lower redshift, even though the mean \index{metallicity} \index{IGM} IGM metallicity 
has moved well past the \index{metallicity!critical} critical transition metallicity. This finding has
\begin{figure}
\begin{center}
\includegraphics[width=10cm]{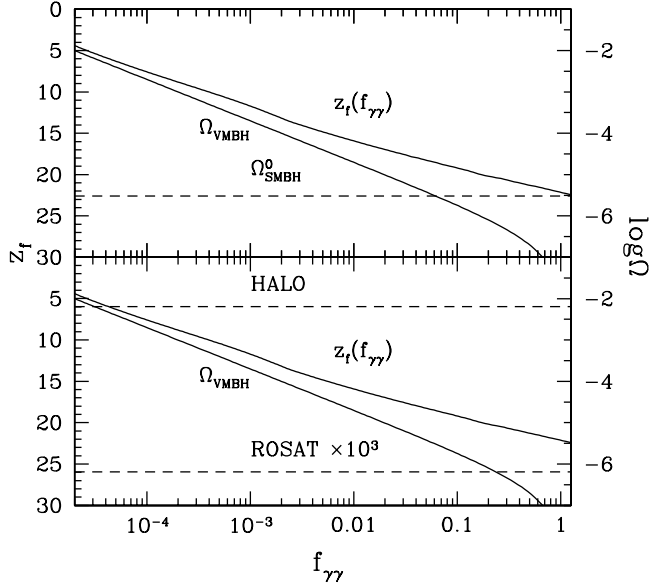}
\end{center}
\caption{Top-heavy to normal IMF 
transition redshift, $z_f$, as a function of \index{supernova!pair-unstable}
SN$_{\gamma\gamma}$ progenitor
mass fraction and the mass density contributed by VMBHs, $\Omega_{VMBH}$.
{\it Top}: The computed critical density of VMBH remnants is compared to the
observed values for SMBHs ({\it upper dashed line}). {\it Bottom}: The
computed critical density of VMBH remnants is compared to the contribution to
$\Omega$ from the \index{X-ray {\it ROSAT} source}  X-ray-bright, off-center {\it ROSAT} sources ({\it lower
dashed line}) and to the abundance predicted assuming that the baryonic dark
matter in galaxy halos is entirely contributed by VMBHs 
({\it upper dashed line}). The observations on $\Omega_{VMBH}$ constrain the 
value of $f_{\gamma\gamma}$. For a given $f_{\gamma\gamma}$, the corresponding
value for the transition redshift can be inferred  by the $z_f$ curve.
This Figure is taken from Schneider et al. (2002).}
\label{fig:schneider2}
\end{figure} 
important implications for the development of efficient strategies for the
detection of \index{Pop III} Pop III stars in primeval galaxies. At any given redshift, a
fraction of the observed objects have a metal content that is low enough to
allow the preferential formation of \index{Pop III} Pop III stars. As metal-free stars are
powerful \index{Ly$\alpha $!forest} Ly$\alpha$ emitters (Tumlinson, Giroux \& Shull 2001; Schaerer 2003),
it is natural to use this indicator as a first step in any search for primordial
objects. Scannapieco et al. (2003) derived the probability that  a given
high-redshift Ly$\alpha$ detection is due to a cluster of \index{Pop III} Pop III stars.
In Figure \ref{fig:Eg} the isocontours in the Ly$\alpha$ 
luminosity-redshift plane are shown, indicating this probability for various 
feedback efficiencies $E_g^{III}$. We see that \index{Pop III} Pop III objects populate a 
well-defined region, whose extent is governed by the feedback strength.
Above the typical flux threshold, \index{Ly$\alpha $!emitter} Ly$\alpha$ emitters are potentially 
detectable at all redshifts beyond 5. Furthermore, the fraction of \index{Pop III} Pop III
objects increases with redshift, independently of $E_g^{III}$. For the 
fiducial case, $E_g^{III}=10^{-3}$, the fraction is only a few percent at
$z=4$ but increases to approximately 15\% by $z=6$. So, the Ly$\alpha$ 
emission from already observed high-$z$ sources can indeed be due to \index{Pop III} Pop III
objects, if such stars were biased toward high masses. Hence collecting large
data samples to increase the statistical leverage may be crucial for 
detecting the elusive first stars.
\begin{figure}
\begin{center}
\includegraphics[width=12cm]{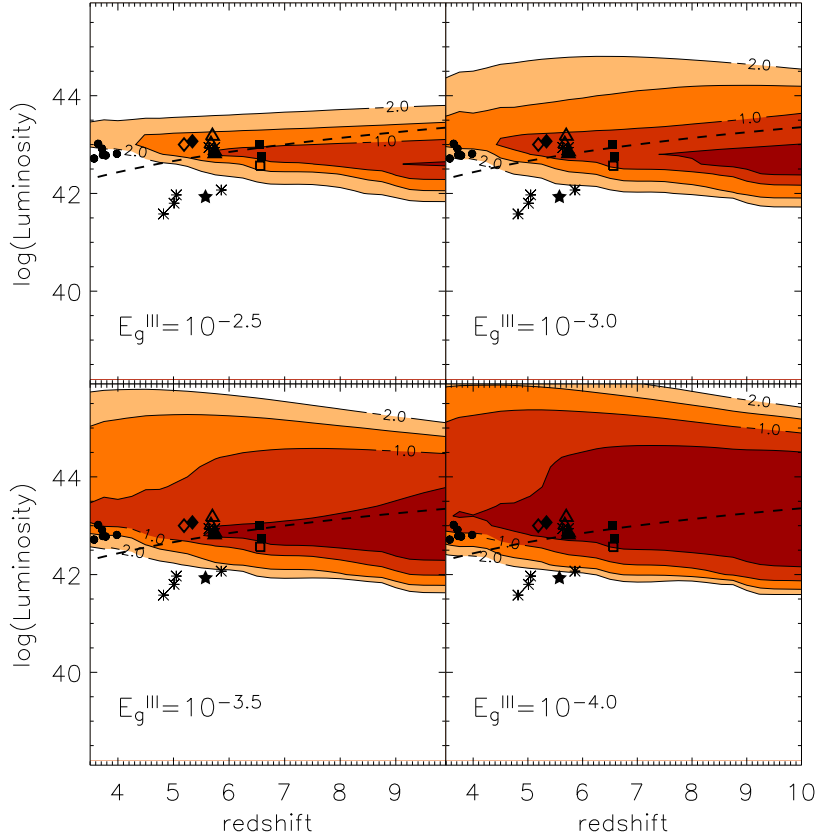}
\end{center}
\caption{Fraction of PopIII objects as a function of Ly$\alpha$
luminosity and redshift.  Isocontours of fractions $\ge 10^{-2},
10^{-1.5}, 10^{-1}$ and $10^{0.5}$ are shown.
Burst-mode star formation with a
$f_\star^{II} = f_\star^{III} = 0.1$ is assumed for all objects.  
In the PopIII case, a lower cutoff mass of $50\Msun$ is assumed.
Each panel is labeled by the assumed $E_g^{III}$ value.
For
reference, the dashed line gives the luminosity corresponding to an
observed flux of $1.5 \times 10^{-17}$ ergs cm$^{-2}$ s$^{-1}$,
and the various points correspond to observed galaxies.
The filled diamond is from Dey et al. (1998),
the filled triangle is from Hu et al. (1999),
the filled star is from Ellis et al. (2001),
the open diamond is from Dawson et al. (2002),
the open square is from Hu et al. (2002),
the asterisks are from Lehnert \& Bremer (2003),
the open triangles are from Rhoads et al. (2003),
the filled circles are from Fujita et al. (2003),
and the filled squares are from Kodaira et al. (2003).
The curves have been extended slightly past $z=4$ for comparison
with the Fujita et al. (2003) data-set. This Figure is taken from
Scannapieco, Schneider \& Ferrara (2003).}
\label{fig:Eg}
\end{figure}

\subsection{\index{feedback!radiative} Radiative Feedback and Reionization}

Before metals are produced, the primary molecule which acquires sufficient
abundance to affect the thermal state of the pristine cosmic gas is molecular
hydrogen, H$_2$. The main reaction able to produce \index{molecular hydrogen} 
molecular hydrogen are 
(Abel et al. 1997)

\begin{eqnarray}\label{equ:H2paths}
\rm
H \  \ \ \  + \ \ e^-  \  & \rightarrow & \ \ {\rm H^-} \ \  +  \ \ h\nu,  \\
\rm H^-  \ \ + \ \ H \ \ & \rightarrow & \ \ \rm H_2  \ \ \  +  \ \  e^-, 
\end{eqnarray}
and
\begin{eqnarray}
\rm
H^+ \ \ + \ \ H  \ \ & \rightarrow & \ \ {\rm H_2^+} \ \  +  \ \ h\nu,  \\
\rm H_2^+  \ \ + \ \ H \ \ & \rightarrow & \ \ \rm H_2   \ \  +  \ \ H^+,
\end{eqnarray}

In objects with baryonic masses $\simgt 3\times10^4\;\Msun$, gravity dominates
and results in the bottom-up hierarchy of structure formation characteristic
of CDM cosmologies; at lower masses, gas pressure delays the collapse. 
The first objects to collapse are those at the mass scale that separates these
two regimes. Such objects have a virial temperature of several hundred degrees
and can fragment into stars only through molecular cooling (Tegmark et al. 
1997). In other words, there are two independent minimum mass thresholds for
star formation: the Jeans mass (related to accretion) and the cooling mass. 
For the very first objects, the cooling threshold is somewhat higher and sets
a lower limit on the halo mass of $\sim 5\times 10^4\;\Msun$ at $z\sim 20$.

\begin{figure}
\begin{center}
\includegraphics[width=12cm]{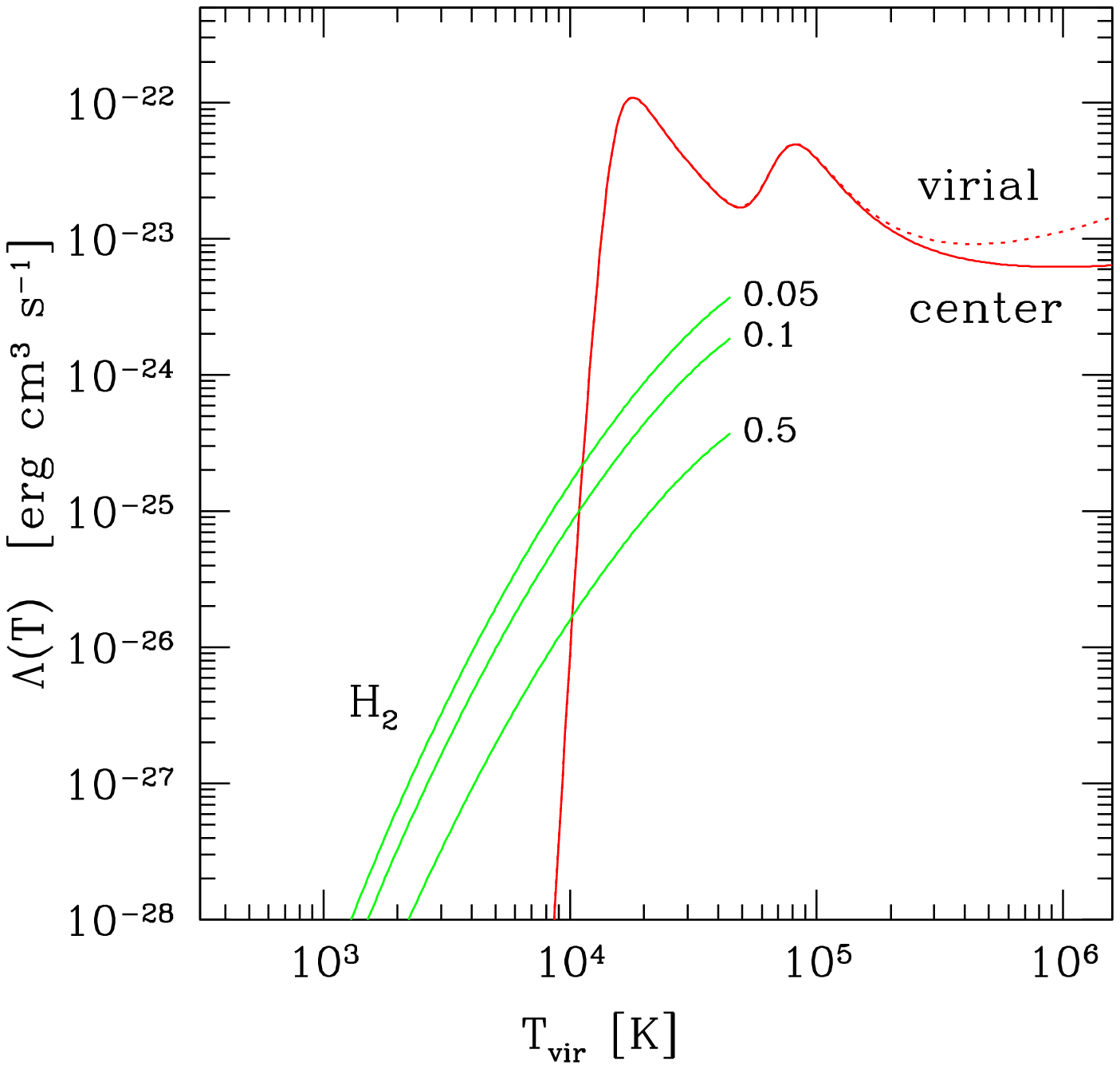}
\end{center}
\caption{Equilibrium cooling curve at the center ({\it solid lines}) and
virial radius ({\it dashed line}) of an isothermal
halo at $z=9$, as a function of virial temperature $T_{vir}$,
and in the absence of a photoionizing background (i.e. prior to the reionization epoch). The halo has an assumed baryonic mass fraction of $\Omega_b=0.019h^{-2}$.
The gas density dependence of the \index{cooling!function} cooling function at high temperatures is due
to Compton cooling off comic microwave background photons. The labeled curves
extending to low temperatures show the contribution due to H$_2$ for three
assumed values of the metagalactic flux in the Lyman-Werner bands, $4\pi J_{\rm LW}$ (in units of $10^{-21}\,{\rm\,erg\,cm^{-2}\,s^{-1}\,Hz^{-1}}$). This Figure is taken from Madau,
Ferrara \& Rees (2001).}
\label{fig:cool}
\end{figure}

\begin{figure}
\begin{center}
\includegraphics[width=12cm]{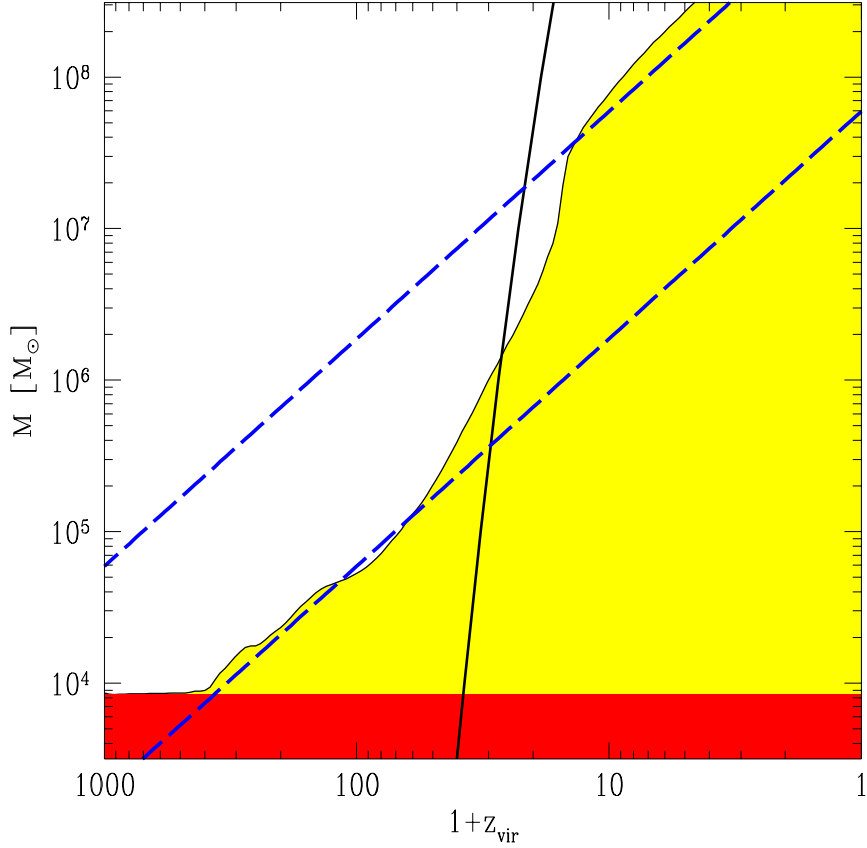}
\end{center}
\caption{The minimum mass needed to collapse.
The function $M_c(z_{vir})$ is plotted as a function on
virialization redshift for standard CDM 
($\Omega=1$, $\Omega_b=0.06$, $h=0.5$).
Only lumps 
whose parameters $(z_{vir},M)$ lie above the shaded area can collapse and
form \index{object!luminous} luminous objects. The dashed straight lines corresponding to $T_{vir}=10^4$ K
and $T_{vir}=10^3$ K are shown for comparison (dashed). 
The dark-shaded region is that in which no radiative 
cooling mechanism whatsoever
could help collapse,
since $T_{vir}$ would be lower than the CMB temperature.
The solid line corresponds to $3-\sigma$ peaks in standard CDM, normalized to 
$\sigma_8=0.7$, so such objects with baryonic mass 
$\Omega_b\times 2\times 10^6\;\Msun \sim 10^5\;\Msun$ can form at $z=30$. 
This figure is taken from Tegmark et al. (1997).}
\label{fig:zm}
\end{figure}

As the first stars form,
their photons in the energy range 11.26-13.6 eV are able to
penetrate the gas and photodissociate H$_{2}$ molecules both in the \index{IGM} IGM
and in the nearest collapsing structures, if they can propagate that far
from their source. Thus, the existence of an \index{UV background} UV
background below the Lyman limit due to \index{Pop III} Pop III objects, capable
of dissociating the H$_{2}$, could deeply influence subsequent 
small structure formation.

Ciardi, Ferrara \& Abel (2000) have shown that the UV flux from these objects
results in a soft (Lyman -- Werner band) \index{UV background!soft} UV background (SUVB), $J_{LW}$, whose
intensity
(and hence \index{feedback!radiative} radiative feedback efficiency) depends on redshift.
At high redshift the radiative feedback can be induced also by the direct
dissociating flux from a nearby object.
In practice, two different situations can occur: i) the collapsing
object is outside the dissociated spheres produced by pre-existing objects:
then its formation could be affected only by the \index{UV background!soft}
SUVB ($J_{LW,b}$), as by construction the direct flux
($J_{LW,d}$) can only dissociate \index{molecular hydrogen} molecular hydrogen inside this region on time
scales shorter than the Hubble time;
ii) the collapsing object is located
inside the dissociation sphere of a previously collapsed object:
the actual dissociating flux in this case is essentially given by
$J_{LW,max}=(J_{LW,b}+J_{LW,d})$.
It is thus assumed that, given a forming Pop~III, if the incident
dissociating flux ($J_{LW,b}$ in the former case, $J_{LW,max}$ in the
latter) is higher than the minimum flux required for negative
feedback ($J_{s}$), the collapse of the object
is halted. This implies the existence of a population of \index{object!dark} "dark objects"
which were not able to produce stars and, hence, light.

To assess the minimum flux required at each redshift to drive the
\index{feedback!radiative} radiative feedback Ciardi et al. (2000) have performed non-equilibrium
multifrequency \index{radiative transfer} radiative transfer calculations for a stellar
spectrum (assuming a metallicity $Z=10^{-4}$) impinging onto a homogeneous gas
layer, and studied the evolution of the following nine species:
H, H$^-$,
H$^+$, He, He$^+$, He$^{++}$, H$_2$, H$_2^+$ and free electrons for
a free fall time. We can then define
the minimum total mass required for an object to self-shield from
an external flux of  intensity $J_{s,0}$ at the Lyman limit, as
$M_{sh}=(4/3) \, \pi \langle\rho_{h}\rangle R_{sh}^3$, where
$\langle\rho_{h}\rangle$ is
the mean dark matter density of the halo in which the gas collapses;
$R_{sh}$ is the shielding radius beyond which \index{molecular hydrogen} 
molecular hydrogen
is not photodissociated and allowing the collapse to take place
on a free-fall time scale.

Values of $M_{sh}$ for different values of $J_{s,0}$ have
been obtained at  various redshifts (see Figure \ref{fig5}).
Protogalaxies with masses above $M_H$ for    which
cooling is predominantly contributed by Ly$\alpha$ line 
are  not affected by the \index{feedback!radiative} radiative feedback studied here.
The collapse of very small objects with mass $<
M_{crit}$ is on the other hand not possible since the
cooling time is longer than the Hubble time.
Thus \index{feedback!radiative} radiative feedback is important 
in the mass range $10^6-10^8 M_\odot$, depending on redshift. 
In order for the negative feedback to be effective,
fluxes of the order of $10^{-24}-10^{-23}$erg s$^{-1}$ cm$^{-2}$
Hz$^{-1}$ sr$^{-1}$ are required.
These fluxes are typically produced by a Pop~III with baryonic mass $10^5
M_{b,5} M_\odot$
at distances closer than $\simeq 21-7\times M_{b,5}^{1/2}$~kpc for the
two above flux values, respectively, while
the \index{UV background!soft} SUVB can reach an intensity in the above range only after $z \approx 15$.
This suggests that at high $z$ negative feedback is driven primarily by the
direct irradiation from neighbor objects in regions of intense clustering,
while only for $z \le 15$ the SUVB becomes dominant.

\begin{figure}
\begin{center}
\includegraphics[width=12cm]{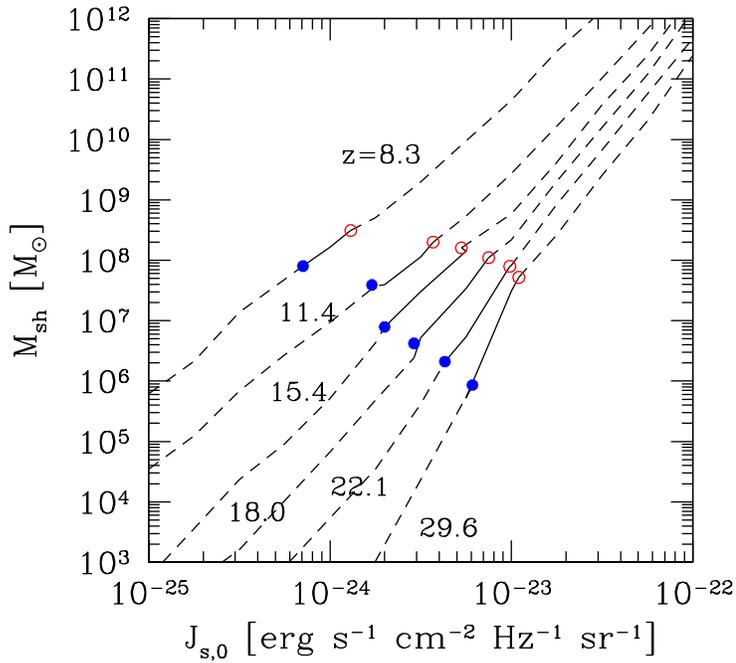}
\end{center}
\caption{Minimum total mass for
self-shielding from an external
incident flux with intensity $J_{s,0}$ at the Lyman limit. The curves
are for different redshift: from the top to the bottom $z=$8.3, 11.4,
15.4, 18.0, 22.1, 29.6. Circles show the value of $M_H$ (open) and
$M_{crit}$ (filled). \index{feedback!radiative} Radiative feedback works only in the solid portions
of the curves. This Figure is taken from Ciardi et al. (2000)}
\label{fig5}
\end{figure}

Figure \ref{fig7} illustrates all possible evolutionary tracks and final fates
of primordial objects, together with the mass scales determined by the various 
physical processes and feedbacks. There are four critical mass
scales in the problem: (i) $M_{crit}$, the minimum mass for an object to be 
able to cool in a Hubble time; (ii) $M_H$, the critical mass for which hydrogen
Ly$\alpha$ line cooling is dominant;
(iii) $M_{sh}$, the characteristic mass 
above which the object is self-shielded, and (iv) 
$M_{by}$ the characteristic mass for \index{feedback!stellar} stellar feedback, below which \index{blowaway} blowaway 
can not be 
avoided. Starting from a virialized dark matter halo, condition (i) produces 
the first branching,  and objects failing to satisfy it will not collapse
and form only a negligible amount of stars. In the following, 
we will refer to these objects as \index{object!dark} {\it dark objects}. 
Protogalaxies with masses in the range $M_{crit} <
M < M_{H}$ are then subject to the effect of \index{feedback!radiative} radiative feedback, 
which could either impede the collapse of those of them with 
mass $M<M_{sh}$, thus contributing to the 
class of \index{object!dark} dark objects, or allow the collapse of the remaining ones ($M>M_{sh}$)
to join those with $M>M_H$ in the class of \index{object!luminous} {\it luminous objects}. This is the 
class of objects that convert a considerable fraction of their baryons in stars.
Stellar \index{feedback!stellar} feedback causes the final bifurcation by inducing a \index{blowaway} blowaway of 
the baryons
contained in \index{object!luminous} luminous objects with mass $M<M_{by}$; this separates the class in
two subclasses, namely "normal" galaxies (although of masses comparable to 
present day
dwarfs) that we dub {\it gaseous galaxies} and tiny stellar aggregates with
negligible traces (if any) of gas to which consequently we will refer to as
{\it naked stellar clusters}. 

\begin{figure}
\begin{center}
\includegraphics[width=12cm]{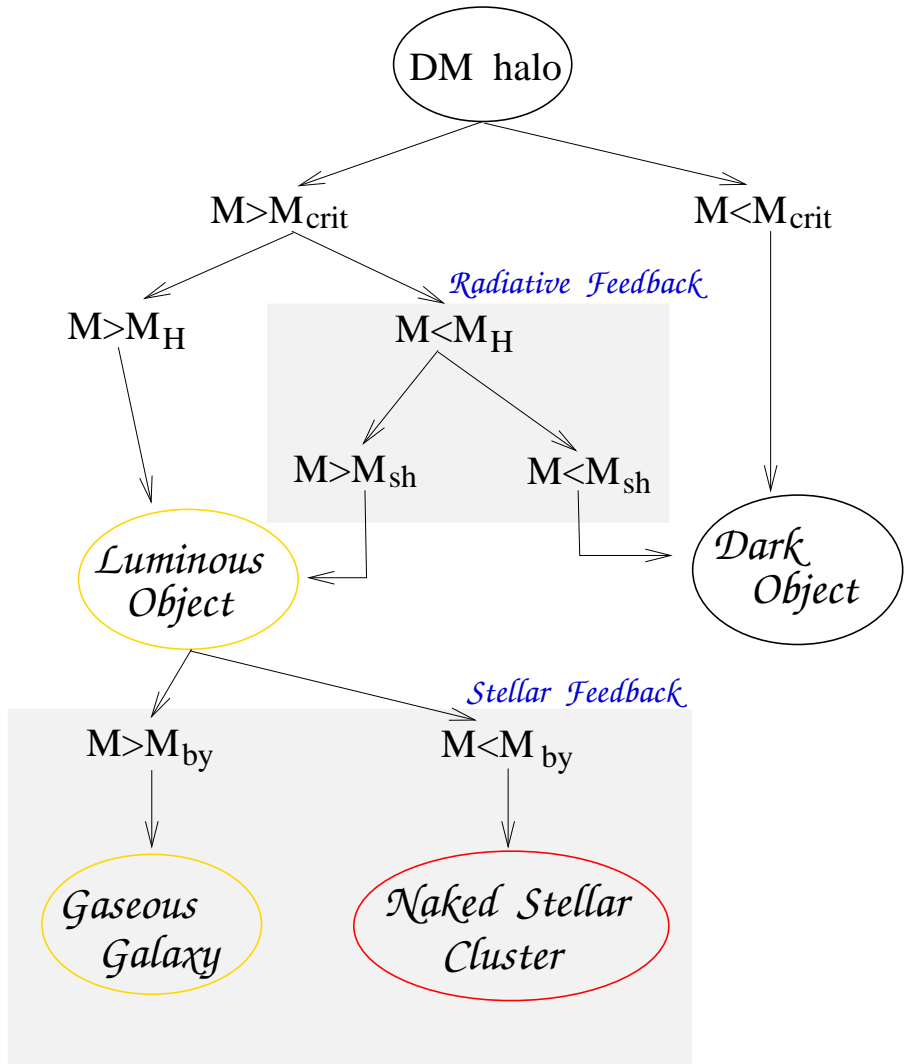}
\end{center}
\caption{Possible evolutionary tracks of objects as
determined by the processes and feedbacks included in the model. See text.
This figure is taken from Ciardi et al. (2000)}
\label{fig7}
\end{figure}

The relative numbers of these objects are shown in Figure~\ref{fig12}.
The straight lines represent, from the top to the bottom,
the number of dark matter halos, \index{object!dark} dark objects, naked
stellar clusters and gaseous galaxies, respectively.
The dotted curve represents the
number of \index{object!luminous} luminous objects with large enough mass
($M>M_H$) to make the H line cooling efficient and become insensitive to
the negative feedback. We remind that the naked stellar
clusters are the \index{object!luminous} luminous objects with $M<M_{by}$, while the
gaseous galaxies are the ones with $M>M_{by}$; thus, the number of
\index{object!luminous} luminous objects present at a certain redshift is given by the sum of
naked stellar clusters and gaseous galaxies.
We first notice that the majority of
the \index{object!luminous} luminous objects that are able to form at high redshift will
experience \index{blowaway} blowaway, becoming naked stellar clusters, while only a minor
fraction, and only at $z\simlt$15, when larger objects start to form,
will survive and become gaseous galaxies.
An always increasing number of \index{object!luminous} luminous objects is forming with
decreasing redshift, until $z\sim 15$, where a flattening is seen. This
is  due to the fact that the dark matter halo mass function is
still dominated by small mass objects, but a large fraction of them cannot
form due to the following combined effects: i) toward lower
redshift
the critical mass for the collapse ($M_{crit}$) increases and fewer
objects satisfy the condition $M>M_{crit}$; ii) the \index{feedback!radiative} radiative feedback
due to either the direct dissociating flux or the \index{UV background!soft} SUVB 
increases at low redshift as the SUVB intensity reaches values
significant for the negative feedback effect.
When the number of \index{object!luminous} luminous objects becomes 
dominated by objects with $M>M_H$, by
$z\sim 10$ the population of \index{object!luminous} luminous objects grows again, basically because their
formation is now unaffected by negative \index{feedback!radiative} radiative feedback. 
A steadily increasing number   of \index{object!dark} objects is prevented from forming stars
and remains dark; this population is about $\sim 99$\% of the total
population of dark matter halos at $z\sim 8$. This is also due to the
combined effect of points i) and ii) mentioned above.
This population of halos which have failed to produce
stars could be identified with the low mass
tail distribution of the dark galaxies that reveal
their presence through gravitational lensing of quasars. 
It has been argued that this
population of dark galaxies outnumbers normal galaxies by a substantial
amount, and Figure~\ref{fig12} supports this view.

\begin{figure}
\begin{center}
\includegraphics[width=12cm]{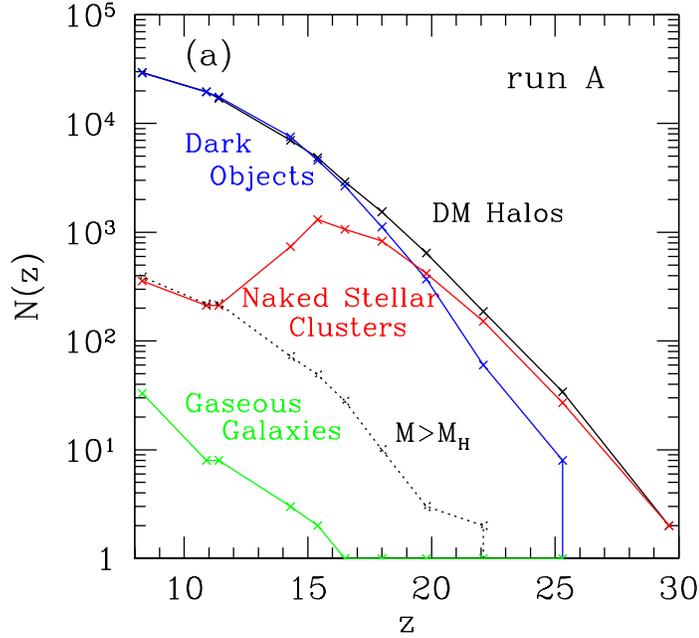}
\end{center}
\caption{Number evolution of different objects in the
simulation box. This Figure is taken from Ciardi et al. (2000)}
\label{fig12}
\end{figure}

\subsubsection{Escape of \index{ionizing photon} Ionizing Photons}

One crucial point to understand cosmic reionization is to determine the
fraction of \index{ionizing photon}ionizing photons that can escape from their parent galaxy.
By solving the time-dependent radiation transfer problem of stellar
radiation through evolving \index{superbubble} superbubbles within a smoothly varying H~I
distribution, Dove et al. (2000) have estimated the fraction of \index{ionizing photon} ionizing 
photons emitted by OB
associations that escapes the H~I disk of our Galaxy into the halo and
intergalactic medium \index{IGM} (IGM).  They considered both coeval star-formation and a
Gaussian star-formation history with a time spread $\sigma_t = 2$ Myr; the
calculations are performed both a uniform H~I distribution and a two-phase 
(cloud/intercloud) model, with a negligible filling factor of hot gas.  
The shells of the expanding superbubbles quickly trap or attenuate the ionizing
flux, so that most of the escaping radiation escapes shortly after the
formation of the \index{superbubble} superbubble. Superbubbles of large associations can \index{blowout} blow
out of the H~I disk and form dynamic chimneys, which allow the ionizing
radiation to escape the H~I disk directly. However, blowout occurs when the
\index{ionizing photon}
ionizing photon luminosity has dropped well below the association's maximum
luminosity.  For the coeval star-formation history, the total fraction of
Lyman Continuum photons that escape both sides of the disk in the solar
vicinity is $\langle f_{esc} \rangle \simeq 0.15 \pm 0.05$ (Figure \ref{fig:dove}). 
For the Gaussian star formation history, $\langle f_{esc} \rangle \simeq 0.06 \pm 0.03$.

\begin{figure}
\begin{center}
\includegraphics[width=12cm]{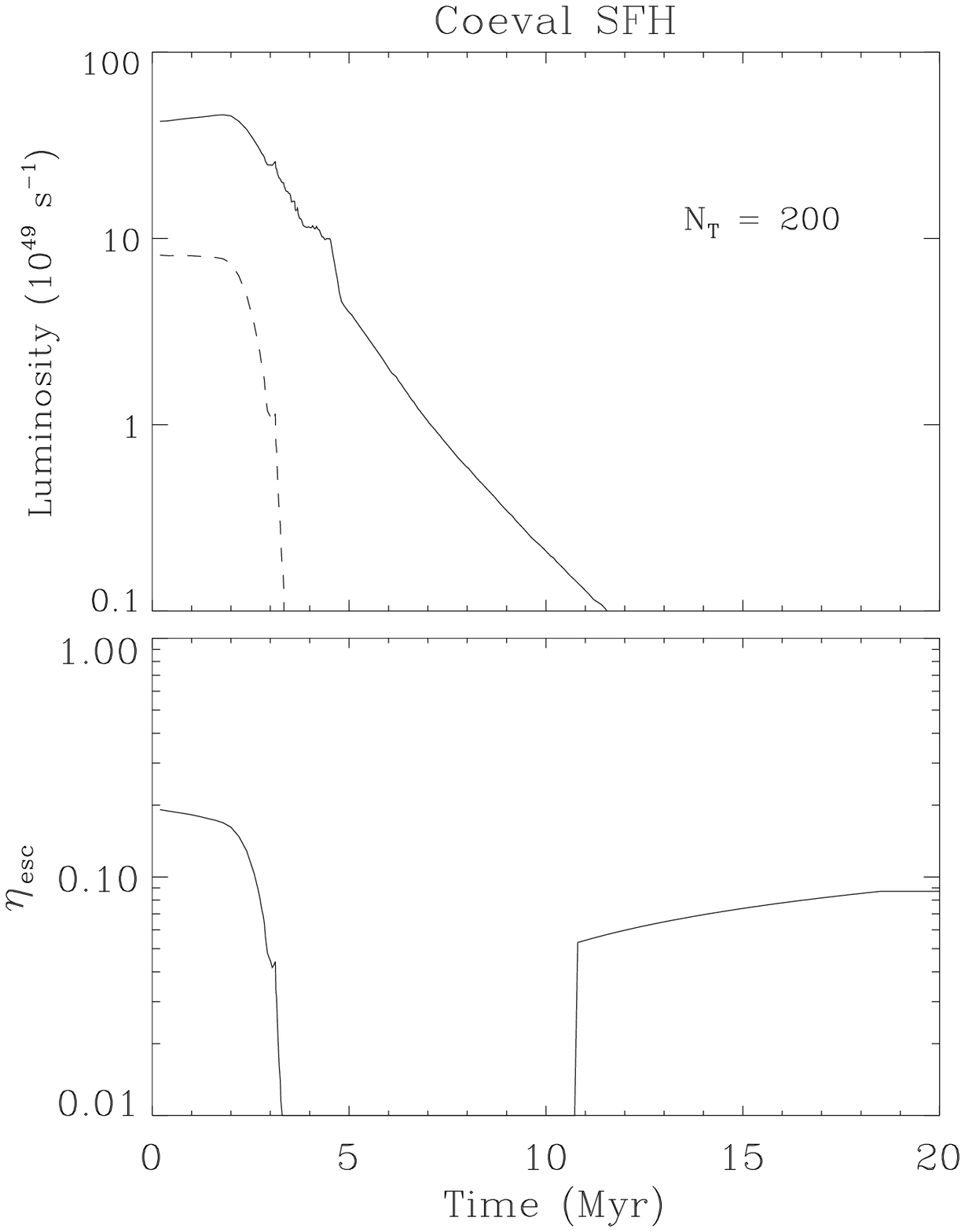}
\end{center}
\caption{Photon luminosity emitted by a single OB association (solid line)
and the luminosity of photons escaping each side of the H~I disk (dashed
line) for coeval star-formation. Also shown is the fraction of photons
emitted that escape, $\eta_{esc}(N_T,t) = S_{esc}(t)/S(t)$. Here, $Z_{tr}=
2\sigma_h$, the cavity height is $Z_{cy} = 4\sigma_h$, and $\beta =1$. The
I-front is trapped within the shell during the time interval where $\eta_{esc}
= 0$. After blowout, $\eta_{esc}$ suddenly increases even though the photon
luminosity of the association is relatively low at these times. Case with
$N_\star=200$. This Figure is taken from Dove et al. (2000)}
\label{fig:dove}
\end{figure}

Bianchi,  Cristiani \& Kim (2001) derived the HI-ionizing background, 
resulting from the integrated contribution of QSOs and galaxies, taking into 
account the opacity of the intervening \index{IGM} IGM.
They have modeled the \index{IGM} IGM with pure-absorbing clouds, with a distribution
in column density of neutral hydrogen, $N_{\rm HI}$, and redshift, $z$,
derived from recent observations of the \index{Ly$\alpha $!forest} Ly$\alpha$ forest
(Kim et al. 2002) and from Lyman Limit systems. The QSOs emissivity
has been derived from the recent fits of Boyle et al. (2000), while
the stellar population synthesis model of Bruzual \& Charlot (2003) and 
a star-formation history from UV observations for the galaxy emissivity. 
Due to the present uncertainties in models
and observations, three values for the fraction of \index{ionizing photon} ionizing
photons that can escape a galaxy interstellar medium, $f_{esc}$=
0.05, 0.1 and 0.4, as suggested by local and high-$z$ UV observations of
galaxies, respectively, are used.

In Figure~\ref{j912} is shown the modeled \index{UV background} UV background, $J(\nu,z)$, at the 
Lyman limit as a function of redshift (solid lines) for the flat universe 
with $\Omega_{m}$=1. The total background is shown as the sum of 
the QSO contribution (the same in each model; dotted line) and the 
galaxy contribution (scaled with $f_{esc}$; dashed lines).

At high redshift, the value of the \index{UV background} UV background is constrained by the
analysis of the \index{effect!proximity} proximity effect, i.e. the decrease in the number of
intervening \index{absorption} absorption lines that is observed in a QSO spectrum when 
approaching the QSO's redshift (Bajtlik et al. 1988). Using high 
resolution spectra,
Giallongo et al. (1996) derived $J(912$\AA$)=5.0_{-1}^{+2.5} 
\times 10^{-22}\;\mathrm{erg\; cm^{-2}\; s^{-1}\; Hz^{-1}\; sr^{-1}}$ for 
$1.7 < z < 4.1$ (see also Giallongo et al. 1999). Larger values are
obtained (same units) by Cooke et al. (1997), $J(912$\AA$)=1.0_{-0.3}^{+0.5} \times
10^{-21}$ for $2.0 < z < 4.5$ and by a recent re-analysis of moderate resolution spectra  by
Scott et al. (2000) who found $J(912$\AA$)=7.0_{-4.4}^{+3.4} 
\cdot 10^{-22}$ in the same redshift range. These measurements are shown in Figure~\ref{j912} with
a shaded area: the spread of the measurements obtained with different
methods and data gives an idea of the uncertainties associated with the
study of the \index{effect!proximity} proximity effect. 
Several models can produce a value of $J(912$\AA$)$ compatible with one
of the measurements from the \index{effect!proximity} proximity effect at $z\sim 3$ shown in  
Figure~\ref{j912}, from a simple QSO-dominated background to models with 
$f_{esc}\sim 0.2$, leading to a limit of $f_{esc}\simlt 20$\%.

\begin{figure}
\begin{center}
\includegraphics[width=11cm]{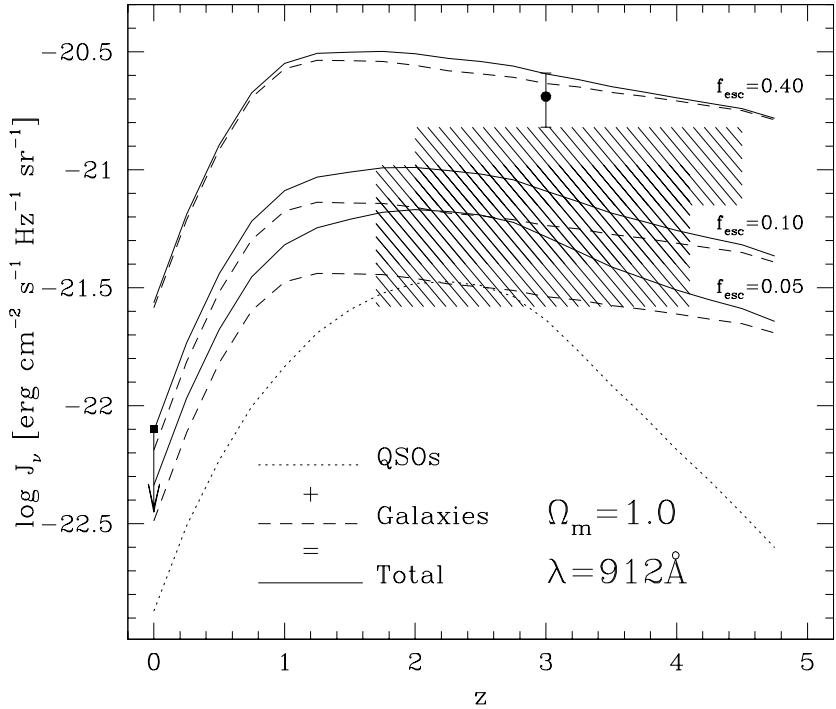}
\end{center}
\caption{UV background at $\lambda=912$\AA\ for the models with
$f_{esc}$=0.05, 0.1  and 0.4 (solid lines), in a flat
$\Omega_{m}=1$ universe. Also shown are the 
separate contribution of the QSOs (dotted line) and of the galaxies
(dashed lines, each corresponding to a value of $f_{esc}$).
The shaded area refer to the Lyman limit UV background estimated from 
the \index{effect!proximity} proximity effect (Giallongo et al. 1996; 
Cooke et al. 1997; Scott et al. 2000). The arrow shows an upper limit for the 
local ionizing background 
(Vogel et al. 1995). The datapoint at $z=3$ is derived from a composite 
spectrum of Lyman-break galaxies (Steidel et al. 2001). The models
and the datapoint of Steidel et al. (2001) have been multiplied by a
$z$-dependent factor, to take into account the cloud emission. This
Figure is taken from Bianchi et al. (2001)}
\label{j912}
\end{figure}

\subsubsection{Cosmological \index{ionization!front} Ionization Fronts}

The \index{radiative transfer} radiative transfer equation in cosmology describes the time evolution of
the specific intensity $I_\nu=I(t,x,\omega,\nu)$ of a diffuse radiation field
(Ciardi, Ferrara \& Abel 2000):

\q
\frac{1}{c}\frac{\partial I_\nu}{\partial t}+\frac{\hat{n}\cdot \nabla I_\nu}{a_{em}}-\frac{H(t)}{c}\left[v\frac{\partial I_\nu}{\partial \nu}-3I_\nu\right]=\eta_\nu -
\chi I_\nu,
\nq

\noindent
where $c$ is the speed of light, $\chi$ is the continuum \index{absorption} absorption coefficient
per unit length along the line of sight, and $\eta_\nu$ is the proper space
averaged volume emissivity. 

\bigskip

If massive stars form in \index{Pop III} Pop III objects, their photons with $h\nu>13.6$~eV
create a cosmological HII region in the surrounding \index{IGM} IGM. Its radius, $R_i$,
can be estimated by solving the following standard equation for the 
evolution of the \index{ionization!front} ionization front:

\q\label{eq:dRi}
\frac{dR_i}{dt}-HR_i=\frac{1}{4\pi n_H R_i^2}\left[S_i(0)-\frac{4}{3}\pi R_i^3
n_H^2 \alpha^{(2)}\right];
\nq

\noindent
note that ionization equilibrium is implicitly assumed. $H$ is the Hubble 
constant, $S_i(0)$ is the \index{ionizing photon} ionizing photon rate, $n_H$ is the \index{IGM} IGM hydrogen
number density and $\alpha^{(2)}$ is the hydrogen recombination rate to 
levels $\geq 2$. For $a R\gg c/H$ the cosmological expansion term $HR_i$ can
be safely neglected. In its full form equation (\ref{eq:dRi}) must be solved
numerically. However, if steady-state is assumed ($dR_i/dt\simeq 0$), then
$R_i$ is approximately equal to the \index{Stromgren radius} Stromgren radius 
$R_s(z)=[(3dN_\gamma/dt)/(4\pi \alpha_2 n_H^2(z))]^{1/3}$:

\q
R_i\simeq R_s=0.05(\Omega_b h^2)^{-2/3}(1+z)_{30}^{-2}S_{47}^{1/3} {\rm kpc},
\nq

\noindent
where $S_{47}=S_i(0)/10^{47} {\rm s^{-1}}$. The approximate expression for
$S_i(0)$ is $\simeq 1.2\times10^{48}f_bM_6$ s$^{-1}$ when a 20\% escape 
fraction for \index{ionizing photon} ionizing photons is assumed and 
$f_b$ is the fraction of baryon that is able to cool and become available to
form stars. In general $R_s$ represents an upper limit for $R_i$, since the
\index{ionization!front} ionization front completely fills the time-varying 
\index{Stromgren radius} Stromgren radius only at
very high redshift, $z\simeq 100$.

\subsubsection{Reionization after \index{{\it WMAP}} {\it WMAP}}
 
The {\it WMAP} satellite has detected an excess in the CMB TE cross-power 
spectrum on large angular scales ($l<7$) indicating an \index{optical depth} optical depth to the
CMB last scattering surface of $\tau_e=0.17\pm0.04$ (Kogut et al. 2003). 
The reionization of the universe must therefore have begun at relatively 
high redshift. Ciardi, Ferrara \& White (2003) have  studied the reionization
process using supercomputer simulations of a large and representative region of
a universe which has cosmological parameters consistent with the {\it WMAP} 
results. The simulation follow both the \index{radiative transfer} radiative transfer of \index{ionizing
photon} ionizing photons
and the formation and evolution of the galaxy population which produces them.

They have shown that the {\it WMAP} measured \index{optical 
depth} optical depth to electron 
scattering is easily reproduced by a model in which reionization is caused by
the first stars in galaxies with total masses of a few $\times10^9\;\Msun$. 
Moreover, the first
stars are `normal objects', i.e. their mass is in the range 1-50 $\Msun$,
but metal-free. Among the different model explored, the `best' {\it WMAP} 
value for $\tau_e$ is matched assuming a \index{IMF!top-heavy} moderately top-heavy IMF (Larson IMF
with $M_c=5\;\Msun$) and an escape fraction of 20\%. A \index{IMF!Salpeter} 
Salpeter IMF with
the same escape fraction gives $\tau_e=0.132$, which is still within the WMAP
68\% confidence range. Decreasing $f_{esc}$ to 5\% gives $\tau_e=0.104$, 
which disagrees with {\it WMAP} only at the 1.0 to 1.5$\sigma$ level. 
In Figure \ref{fig:ciardi1} is shown the neutral hydrogen number density at
three redshift ($z=17.6$, 15.5, and 13.7) and in Figure \ref{fig:ciardi3} is
shown the evolution of the electron \index{optical depth} optical depth for the different IMFs 
considered.

In the best-fit model reionization is essentially complete by $z_r\simeq 13$. 
This is difficult to reconcile with observation of the \index{effect!Gunn-Peterson} Gunn-Peterson effect
in $z>6$ quasars (Becker et al. 2001; Fan et al. 2002) which imply a 
volume-averaged neutral fraction $ \simgt 10^{-3}$ and a mass-averaged neutral
fraction $\sim 1$\% at $z=6$. Then a fashinating (although speculative) possibility
is that the universe was reionized twice (Cen 2003; Whyte \& Loeb 2002) with
a relatively short redshift interval in which the \index{IGM} IGM became neutral again.

\begin{figure}
\begin{center}
\includegraphics[width=12cm]{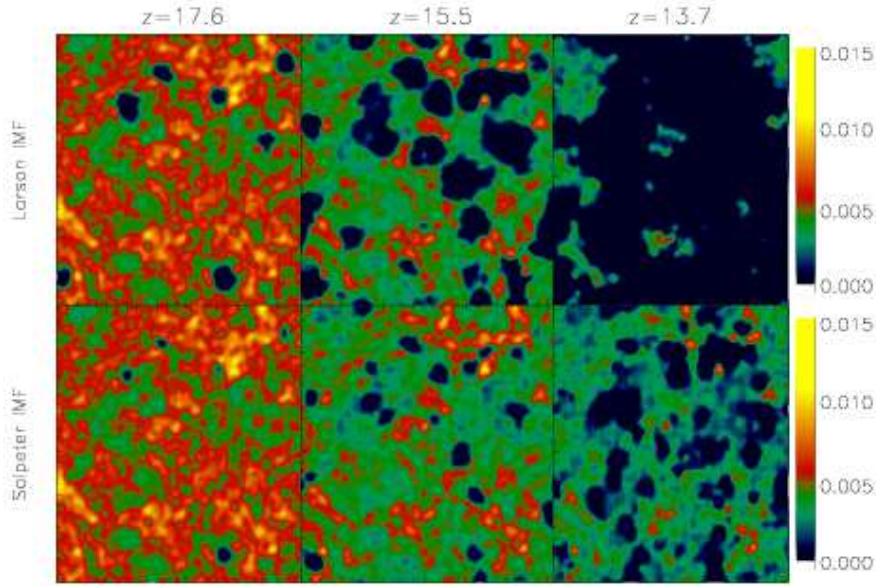}
\end{center}
\caption{Slice through the simulation boxes.
The six panel show the neutral hydrogen number density for the case of a 
Larson IMF with $M_c=5\;\Msun$ (upper panels) and for a Salpeter IMF (lower
panels) at redshifts, from left to right, $z=17.6,\;15.5$ and $13.7$. 
The escape fraction is 20\%. The box for the \index{radiative transfer} radiative transfer simulation 
has a comoving length of $L=20h^{-1}$ Mpc. This Figure is taken from Ciardi et
al. (2003).}
\label{fig:ciardi1}
\end{figure}

\begin{figure}
\begin{center}
\includegraphics[width=12cm]{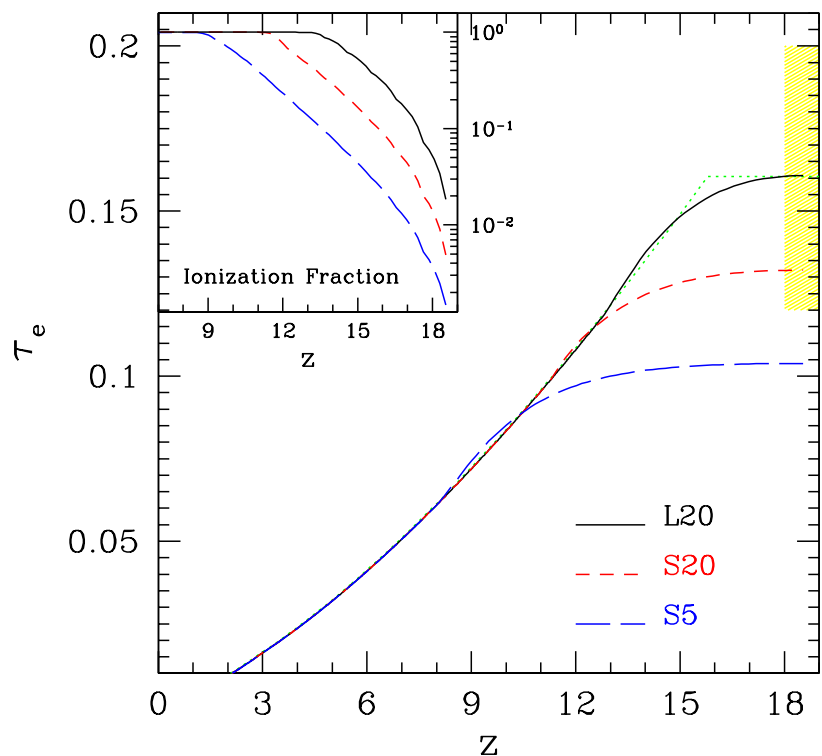}
\end{center}
\caption{Redshift evolution of the electron 
optical depth, $\tau_e$, for an Salpeter IMF with $f_{esc}=5$\% (long-dashed
line) and $f_{esc}=20$\% (short-dashed line), and for a Larson IMF with 
$M_c=5\;\Msun$ and $f_{esc}=20$\%. The dotted line refers to a sudden 
reionization at $z=16$. The shaded region indicates the optical depth 
$\tau_e=0.16\pm0.04$ (68\% CL) implied by the Kogut et al. (2003) `model
independent' analysis. In the inset the redshift evolution of the 
volume-averaged ionization fraction, $x_v$, is shown for the three runs.
This Figure is taken from Ciardi et al. (2003).}
\label{fig:ciardi3}
\end{figure}

\clearpage
\newpage

\begin{thereferences} 
\item Abel T., Anninos P., Zhang Y. \& Norman M. L., 1997, {\it NewA}, {\bf 2}, 181
\item Abel T., Bryan G. L., Norman M. L. 2000, {\it ApJ},
 {\bf 540}, 39
\item Balbus S. A., 1995, {\it The Physics of the Interstellar Medium and 
Intergalactic Medium}, ASP Series Vol. 80 (San Francisco: PASP) p. 328 
\item  Bajtlik S., Duncan R. C. \& Ostiker J. P., 1998, {\it ApJ}, {\bf 327}, 570
\item  Becker R. H. et al., 2001, {\it AJ}, {\bf 122}, 2850
\item  Benson A. J., Lacey C. G., Bauch C. M., Cole S., Frenk C. S., 
2002, {\it MNRAS}, {\bf 333}, 156
\item  Benson A. J. \& Madau P., 2003, {\it MNRAS}, {\bf 344}, 835
\item  Bianchi S., Cristiani S. \& Kim T.-S., 2001, {\it A\&A}, {\bf 376}, 1
\item  Boyle B. J., Shanks T., Croom S. M., Smith R. J., Miller L., 
Loaring N., Heymans C., 2000, {\it MNRAS}, {\bf 317}, 1014
\item  Bromm V., Coppi P. S., Larson R. B. 1999, {\it ApJ}, {\bf 527}, L5
\item  Bromm V., Coppi P. S., Larson R. B. 2002, {\it ApJ}, {\bf 564}, 23
\item  Bromm V., Ferrara A., Coppi P. S., Larson R. B., 2001, {\it MNRAS}, 
{\bf 328}, 969
\item  Bruzual A. G. \& Charlot S., 2003, {\it MNRAS}, {\bf 344}, 1000 
\item  Cen R., 2003, {\it ApJ}, {\bf 591}, 12
\item  Ciardi B., Ferrara A. \& Abel T., 2000, {\it ApJ}, {\bf 533}, 594
\item  Ciardi B., Ferrara A., Governato F. \& Jenkins A., 2000, {\it MNRAS}, 
{\bf 314}, 611
\item  Ciardi B., Ferrara A., Marri S. \& Raimondo G., 2001, {\it MNRAS}, {\bf 324}, 381 
\item  Ciardi B., Ferrara A. \& White S. M. D., 2003, {\it MNRAS}, {\bf 344}, L7
astro-ph/0302451
\item  Cioffi D., McKee C. F. \& Bertschiger E., 1988, {\it ApJ}, {\bf 334}, 252
\item  Cooke A. J., Espey B. \& Carswell R. F., 1997, {\it MNRAS}, {\bf 284}, 552
\item  Dawson S., Spinrad H., Stern D., Dey A., van Breugel W., de Vries
W., \& Reuland M., 2002, {\it ApJ}, {\bf 570}, 92
\item  Dekel A. \& Silk J., 1986, {\it ApJ}, {\bf 303}, 39
\item  Dey A., Spinrad H., Stern D., Graham J. R., \& Chaffee F. H.,
1998, {\it ApJ}, {\bf 498}, 93
\item  Dove J. B., Shull J. M. \& Ferrara A., 2000, {\it ApJ}, {\bf 531}, 846
\item  Ellis R., Santos M. R., Kneib J.-P., Kuijken K., 2001, {\it ApJL}, 
{\bf 560}, 119
\item  Fan X. et al., 2002, {\it AJ}, {\bf 123}, 1247
\item  Ferrara A. \& Tolstoy E., 2000, {\it MNRAS}, {\bf 313}, 291
\item  Field G. B., 1995, {\it ApJ}, {\bf 142}, 531
\item  Fryer G. M., 1999, {\it ApJ}, {\bf 522}, 413
\item  Fryer G. M., Woosley S. E. \& Weaver T. A., 2001, {\it ApJ}, {\bf 550}, 372
\item  Fujita S. S. et al. 2003, {\it AJ}, {\bf 125}, 13
\item  Furlanetto S. R. \& Loeb A., 2003, {\it  ApJ}, {\bf  588}, 18
\item  Giallongo E., Cristiani S., D'Odorico S., Fontana A. \& Savaglio
S., 1996, {\it  ApJ}, {\bf  466}, 46
\item  Giallongo E., Fontana A., Cristiani S. \& D'Odorico S., 1999, 
{\it ApJ}, {\bf  510}, 605
\item  Haardt \& Madau, 1996, {\it ApJ}, {\bf 461}, 20
\item  Heger A. \& Woosley S. E., 2001, {\it  ApJ}, {\bf  567}, 532
\item  Hu E. M., Cowie L. L., McMahon R. G., Capak P., Iwamuro F., Kneib
J.-P., Maihara T., \& Motohara K., 2002, {\it  ApJ}, {\bf 568}, L75
\item  Hu E. M., McMahon, R. G., Cowie, L. L., 1999, {\it ApJ}, {\bf 522}, L9
\item  Kim T.-S., Cristiani S. \& D'Odorico S., 2002, {\it  A\&A}, {\bf 383}, 747
\item  Kodaira K. et al. 2003,{\it PASJ}, {\bf 55}, 17
\item  Kogut A. et al., 2003, {\it ApJS}, {\bf 148}, 161
\item  Larson R. B., 1974, {\it  MNRAS}, {\bf  169}, 229
\item  Lehnert M. D. \& Bremer M., 2003, {\it ApJ}, {\bf 593}, 630
\item  Mac Low M.-M. \& Ferrara A., 1999, {\it  ApJ}, {\bf  513}, 142
\item  Madau P., Ferrara A. \& Rees M., 2001, {\it  ApJ}, {\bf  555}, 92
\item  Madau P. \& Rees M. J., 2001, {\it  ApJ}, {\bf  551}, L27
\item  Marri S. et al. 2003, in preparation
\item  Marri S. \& White S. D. M., 2003, {\it MNRAS}, {\bf 345}, 561
\item  McKee C. F. \& Draine B. T., 1991, {\it  Science}, {\bf  252}, 397
\item  Mori M., Ferrara A. \& Madau P., 2002, {\it  ApJ}, {\bf  571}, 40
\item  Navarro J. F. \& White S. D. M., 1993, {\it  MNRAS}, {\bf  265}, 271
\item  Norberg P. et al., 2002, {\it  MNRAS}, {\bf 336}, 907
\item  Omukai K., 2001, {\it  ApJ}, {\bf  546}, 635
\item  Omukai K. \& Inutsuka S. 2002, {\it MNRAS}, {\bf 332}, 59
\item  Omukai K. \& Palla F. 2003, {\it ApJ}, {\bf 589}, 677
\item  Osterbrok D. E., 1989, {\it Astrophysics of Gaseous Nebulae and Active
Galactic Nuclei}, (Mill Valley: University Science Books)
\item  Ostriker J. P. \& McKee C. F., 1988, {\it  Rev. Mod. Phys.}, {\bf  60}, 1
\item  Pearce F. R. et al., 1999, {\it  ApJ}, {\bf  521}, 99
\item  Persic M., Salucci P., Stel F., 1996, {\it MNRAS}, {\bf 281}, 27
\item  Ricotti M., Ferrara A. \& Miniati F., 1997, {\it  ApJ}, {\bf  485}, 254
\item  Rhoads J. E. et al., 2003, {\it  AJ}, {\bf  125}, 1006 
\item  Scannapieco E., Ferrara A. \& Broadhurst T., 2000, {\it  ApJL}, {\bf 536}, 11
\item  Scannapieco E., Ferrara A. \& Madau P., 2002, {\it  ApJ}, {\bf  574}, 590 
\item  Scannapieco E., Schneider R. \& Ferrara A., 2003, {\it  ApJ}, {\bf  589}, 35
\item  Schaerer D., 2002, {\it  A\&A}, {\bf  382}, 28
\item  Schaye J., Rauch M., Sargent W. L. W. \& Kim T.-S., 2002, {\it  ApJL},
{\bf 541}, 1
\item  Schneider R., Ferrara A., Natarayan P. \& Omukai K., 2002, {\it  ApJ},
{\bf 571}, 30
\item  Schneider R., Ferrara A., Salvaterra R., Omukai K. \& Bromm V., 
2003, {\it  Nature}, {\bf  422}, 869
\item  Scott J., Bechtold J., Dobrzycki A. \& Kulkarni V. P., 2000, {\it  ApJS}, {\bf 130}, 67
\item  Sedov L., 1978, {\it Similitude et Dimensions en Mecanique}, (Moscow: 
Editions de Moscou)
\item  Spizer L., 1978, {\it Physical Processes in the Interstellar Medium},
(Whiley: New York)
\item  Silk J., 2003, {\it MNRAS}, {\bf 343}, 249
\item  Steidel C. C., Pettini M. \& Adelberger K. L., 2001, {\it  ApJ}, {\bf  546}, 665
\item  Tegmark M. et al., 1997, {\it ApJ}, {\bf  474}, 1
\item  Thacker R. J. \& Couchman H. P., 2001, {\it  ApJ}, {\bf  555}, L17
\item  Thomas P. A., Fabian A. C., Arnaud K. A., Forman W., Jones C.,
1986, {\it  MNRAS}, {\bf  222}, 655
\item  Tumlinson J., Giroux M. \& Shull J. M., 2001, {\it  ApJ}, {\bf  555}, 839
\item  Umeda H. \& Nomoto K., 2002, {\it  ApJ}, {\bf  565}, 385
\item  White S. D. M. \& Frenk C. S., 1991, {\it  ApJ}, {\bf  379}, 52
\item  Wyithe S. \& Loeb A., 2003, {\it ApJ}, {\bf 588}, 69
\item  Wolfire M. G., Hollenbach D., McKee C. F., Tielens A. G. G. M.
\& Bakes E. L. O., 1995, {\it  ApJ}, {\bf  443}, 152
\item  
\item  

\end{thereferences} 

\clearpage
\newpage

\input sigrav.ind

\end{document}